\documentclass[twocolumn,showpacs,preprintnumbers,nofootinbib]{revtex4-1}
\usepackage{graphicx,amsmath,amssymb,xcolor,hyperref}
\usepackage[utf8]{inputenc}
\usepackage{ulem}

\def\MET{\rlap{\hspace{0.08cm}/}{E}_T}

\def\dslash{\rlap{\hspace{0.03cm}/}{\partial}}

\def\eDMEFT{e\textsc{DMeft}}

\newcommand{\SU}{\mathrm{SU}}
\newcommand{\vel}{{\rm v}}

\begin{document}

\author{Tommi Alanne}
\email{tommi.alanne@mpi-hd.mpg.de}
\author{Florian Goertz}
\email{florian.goertz@mpi-hd.mpg.de}
\affiliation{Max-Planck-Institut f{\"u}r Kernphysik, Saupfercheckweg 1, 69117 Heidelberg, Germany}
\title{Extended Dark Matter EFT}


\begin{abstract}
Conventional approaches to describe dark matter phenomenology at collider 
and (in)direct detection experiments in the form of dark 
matter effective field theory or simplified models
suffer in general from drawbacks regarding validity at high
energies and/or generality, limiting their applicability. 
In order to avoid these shortcomings, we propose a hybrid framework
in the form of an effective theory, including, however, both the dark matter
states and a mediator connecting the former to the Standard Model fields.
Since the mediation can be realized through rather light new dynamical 
fields 
allowing for non-negligible collider signals in missing 
energy searches,
the framework remains valid for 
the phenomenologically interesting parameter region, while 
retaining correlations dictated by gauge symmetry. 
Moreover, a richer new-physics sector can be consistently
included via higher-dimensional operators.
Interestingly, for fermionic and scalar dark matter with 
a (pseudo-)scalar mediator, the leading effects originate 
from dimension-five operators, allowing to capture them
with a rather small set of new couplings.
We finally examine the correlations between constraints from
reproducing the correct relic density, direct-detection 
experiments, and mono-jet and Higgs\,+\,missing energy signatures 
at the LHC.
\end{abstract}

\maketitle

\section{Introduction and Setup}
\label{sec:intro}

The origin of the dark matter (DM) observed in the universe is one
of the biggest mysteries in physics. A multitude of experiments, which 
are probing very diverse energies, are 
currently running or in preparation to address this question. Experiments aiming for a
direct detection of DM particles via nuclear recoil typically probe
collision energies in the keV range, while collider experiments, trying to
produce DM particles, feature momentum transfers
exceeding the TeV scale.

Combining the results from all kinds of experiments in a single, consistent,
yet general framework is important in order to resolve the nature of DM.
Bounds from direct detection experiments are usually interpreted
in an
effective field theory (EFT) approach, removing the mediator 
that couples the DM particles to the Standard Model~(SM) as an explicit dynamical degree of freedom at low energies. 
This is possible, if the mediator is assumed to be much heavier than the scale of the
experiments, and its effects can thus be described by generic EFT operators, consisting only 
of the SM and DM fields.

On the other hand, collider experiments such as the LHC run at much larger energies,
and it turns out that the momentum transfers lie typically at
(or above) interesting mass ranges for the mediator, to which the analyses are sensitive 
(unless the model is very strongly coupled~\cite{Bruggisser:2016nzw}). 
Thus,
the EFT description becomes invalid since the mediator is missing
in the spectrum~\cite{Busoni:2013lha,Bauer:2016pug}.
In consequence, collider searches are typically interpreted in terms of simplified models
where the mediator is not removed, and its interactions with the SM and the DM
are simply parametrized by $D\leq4$ operators. For the sake
of flexibility, the simplest implementations
do not require gauge-invariance and are thus not well behaved at large energies. 
A  further drawback
of this approach is that they are still specific models that do not allow for a maximally general description of the dark
sector. In the end, one suffers either from a lack of generality or, even worse,
from a lack of validity.

In this article, we want to propose a hybrid framework, which 
can alleviate the above problems:
We will consider the minimal amount of additional {\it dynamical} degrees of freedom---a DM particle {\it and} 
a mediator---which is able to generate the correct DM abundance and allows for testability at TeV scale energies, while retaining
the generality and consistency of the EFT framework.

We will, thus, consider the SM extended by a
SM-singlet particle,
${\cal D}$, that is stable on cosmological scales, and a mediator, ${\cal M}$, that couples it to the SM,
as well as higher-dimensional EFT operators, consisting of these fields.
We will start by assuming the DM to be a fermion, ${\cal D}=\chi$, and 
the mediator a (pseudo)scalar, ${\cal M} = {\cal S} (\tilde{\cal S})$, and then move on to
consider also the case of scalar DM.
In these setups, the leading EFT effects will be at the level of $D=5$ operators, and their
inclusion allows to parametrize physics of the dark sector beyond the single DM
particle and the mediator.\footnote{The generalization
to (fermionic or scalar) DM with a vector mediator calls for the inclusion of
$D=6$ operators and a study of the very
rich phenomenology, 
and we will leave this for future work.}

In fact, while the inclusion of the latter particles
makes the theory valid at collider energies, the augmentation with $D=5$ operators
accounts for the fact that the dark/new sector is likely to be non-minimal.
Indeed, there is no stringent reason to believe that the sector
related with the DM consists only of very few particles, 
while the SM has a very rich structure.
On the other hand, it is conceivable that a few of the new particles (the DM
particle and the mediator) are considerably lighter than the rest of the new physics (NP).
For example, the mediator could be a (pseudo-)Goldstone boson of a spontaneously broken
global symmetry.
The goal of this article is to povide the theoretical framework of this 
extended DM EFT (\eDMEFT) approach,
demonstrating its strength in phenomenological analyses,
as well as pointing out emerging synergies and generic correlations between
observables, which are retained in the EFT approach.

    \subsection{Fermionic DM with a scalar or pseudoscalar mediator}

    The effective Lagrangian of the model described above, with 
    a fermion singlet, $\chi$, and  a (CP even) scalar, ${\cal S}$,
    including operators up to $D=5$ 
    (following normalisations of Ref.~\cite{Carmona:2016qgo}) reads\footnote{The EFT of the SM plus just a
    scalar singlet ${\cal S}$ has been explored in Refs~\cite{Franceschini:2016gxv,Gripaios:2016xuo,Carmona:2016qgo}.} 
    \begin{eqnarray}
    \label{eq:LEFT}
    \mathcal{L}_{\rm eff}^{{\cal S} \chi}& = & {\cal L}_{\rm SM} +
    \frac{1}{2}\partial_\mu {\cal S} \partial^\mu {\cal S} 
    - \frac 1 2 \mu_S^2 {\cal S}^2 + \bar \chi i \dslash\chi- m_\chi \bar \chi \chi \nonumber \\
    &-& \lambda_{S1}^{\prime} v^3 {\cal S}-\frac{\lambda^\prime_{S}}{2 \sqrt 2} v {\cal{S}}^3 - \frac{\lambda_{S}}{4} {\cal{S}}^4 \nonumber \\
    &-& \lambda^\prime_{HS} v |H|^2 {\cal{S}} 
	- \lambda_{HS} |H|^2 {\cal{S}}^2 \nonumber \\[1.5mm]
    &-& y_S {\cal S}  
	\bar \chi_L \chi_R +\mathrm{h.c.}
	\nonumber \\[2.5mm]
    &-& \frac{\cal{S}}{\Lambda} \left[ c_{\lambda S} {\cal{S}}^4 
    +c_{HS} |H|^2 {\cal{S}}^2 
    +c_{\lambda H} |H|^4 \right] \\[1mm]
    &-& \frac{\cal{S}}{\Lambda} \left[(y_d^S)^{ij} \bar{Q}_{\mathrm{L}}^i H d_{\mathrm{R}}^j 
	+ (y_u^S)^{ij}\bar{Q}_{\mathrm{L}}^i\tilde{H}u_{\mathrm{R}}^j \right.\nonumber\\
    && \quad \ + \left.(y_\ell^S)^{ij} \bar{L}_{\mathrm{L}}^i H \ell_{\mathrm{R}}^j +\mathrm{h.c.}\right]\nonumber\\[1mm]
    &-&\frac{y_S^{(2)} {\cal S}^2 + y_H^{(2)} |H|^2 
	}{\Lambda} 
	\bar{\chi}_L \chi_R +\mathrm{h.c.}
	\nonumber\\[1mm]
    &-&\frac{\cal{S}}{\Lambda}\frac{1}{16\pi^2}\left[g^{\prime 2} c_B^S B_{\mu\nu} B^{\mu\nu}+ g^2 c_W^S W^{I\mu\nu} W_{\mu\nu}^I\right.\nonumber\\
    && \qquad \qquad +\left.g_s^2 c_G^S G^{a\mu\nu}G_{\mu\nu}^a\right]\,.\nonumber
    \end{eqnarray}
    Here $Q_{\mathrm{L}}^i$ and $L_{\mathrm{L}}^i$ are the $i$-th generation left-handed $\SU(2)_{\mathrm{L}}$ quark and lepton
    doublets, resp., $d^j_{\mathrm{R}}$, $u^j_{\mathrm{R}}$, and $\ell_{\mathrm{R}}^j$ are the right-handed 
    singlets for generation $j$, and $H$ is the Higgs doublet. 
    The Higgs doublet develops a vacuum expectation value (vev),  $|\langle H \rangle| \equiv  v/\sqrt 2 \simeq 174$~GeV, triggering
    electroweak symmetry breaking (EWSB).  In unitary gauge, the Higgs field is expanded around the vev as 
    $H \simeq 1/\sqrt2 (0, v + h)^T$. Here, $h$ is the physical Higgs boson, with mass
    $m_h \approx 125$\,GeV. We assume that the mediator does not develop a vev and have, thus, included a linear term in $\cal S$.

    Besides the SM couplings, there are several new interactions,
    both at the renormalizable $(D=4)$ level and in the form of effective 
    $D=5$ operators.
    In the scalar sector, the cubic and quartic
    terms in the singlet potential are parametrized by the couplings 
    $\lambda^\prime_{S}$ and $\lambda_{S}$, resp., while the Higgs-portal couplings
    involving one or two singlets are denoted by  
    $\lambda^\prime_{HS}$ and $\lambda_{HS}$,
    resp. Note that, after EWSB, the latter coupling provides a contribution to
    the mediator mass, which is given by
    \begin{equation}
    m_S = \sqrt{\mu_S^2 + \lambda_{HS} v^2}\,.
    \end{equation}
    In addition, there is a Yukawa coupling between the scalar mediator,
    $\cal S$, and the DM fermions, $\chi$, denoted by $y_S$.
    At the $D=5$ level, all interactions are suppressed by one power of
    the scale of heavy NP, $\Lambda$, which mediates contact interactions 
    between the various fields.
    In the pure scalar sector, gauge invariant terms feature four ($c_{\lambda H}$), two
    ($c_{HS}$), or zero ($c_{\lambda S}$) Higgs fields. In order to couple
    the mediator, $\cal S$, to SM fermions, the presence of a Higgs doublet is required,
    allowing for $D=5$ Yukawa-like couplings, $\sim (y^S_d)^{ij}\!,(y^S_u)^{ij}$.
    Scalar couplings to the DM fermions at the $D=5$ level, on the other hand, 
    involve either two scalar singlets or two doublets, due to gauge invariance,
    parametrized by $y_S^{(2)}$ and $y_H^{(2)}$, respectively.
    Finally, there are effective couplings of $\mathcal{S}$ to the $\mathrm{U}(1)_Y$, $\SU(2)_{\mathrm{L}}$,
    and $\SU(3)_{\mathrm{c}}$ field strengths squared, denoted by $c_B^S$, $c_W^S$, 
    and $c_G^S$. In the following, we assume the interactions with ${\cal S}$ to conserve CP, and thus
    all coefficients in the Lagrangian of Eq.~\eqref{eq:LEFT}  are real. 
    
    We conclude this discussion noting that, if the new sector residing at the scale $\Lambda$ is governed
    by a coupling $g_\ast$, the effective coefficients above can be assigned a certain scaling in this coupling (see e.g.
    \cite{Luty:1997fk,Cohen:1997rt,Giudice:2007fh,Chala:2017sjk}), which in our case (without assuming the Higgs
    to be a pseudo-Goldstone state) reads $c_{\lambda S} \sim c_{H S} 
    \sim c_{\lambda H} \sim g_\ast^3\,,\ y_f^S \sim y_f g_\ast\,,\ y_{S,H}^{(2)} \sim g_\ast^2\,, c_V^S \sim g_\ast$.
    This allows to order the operators according to their expected importance in a certain coupling regime and
    can for example be used to reduce the number of $D=6$ operators to be considered to leading approximation
    in the case of a vector mediator.\footnote{As shown recently \cite{Goertz:2017gor}, such a counting could also 
    allow to lift the ambiguity in determining masses of new states in an EFT approach.}

\begin{figure}[!t]
    \begin{center}
	\includegraphics[height=2.6in]{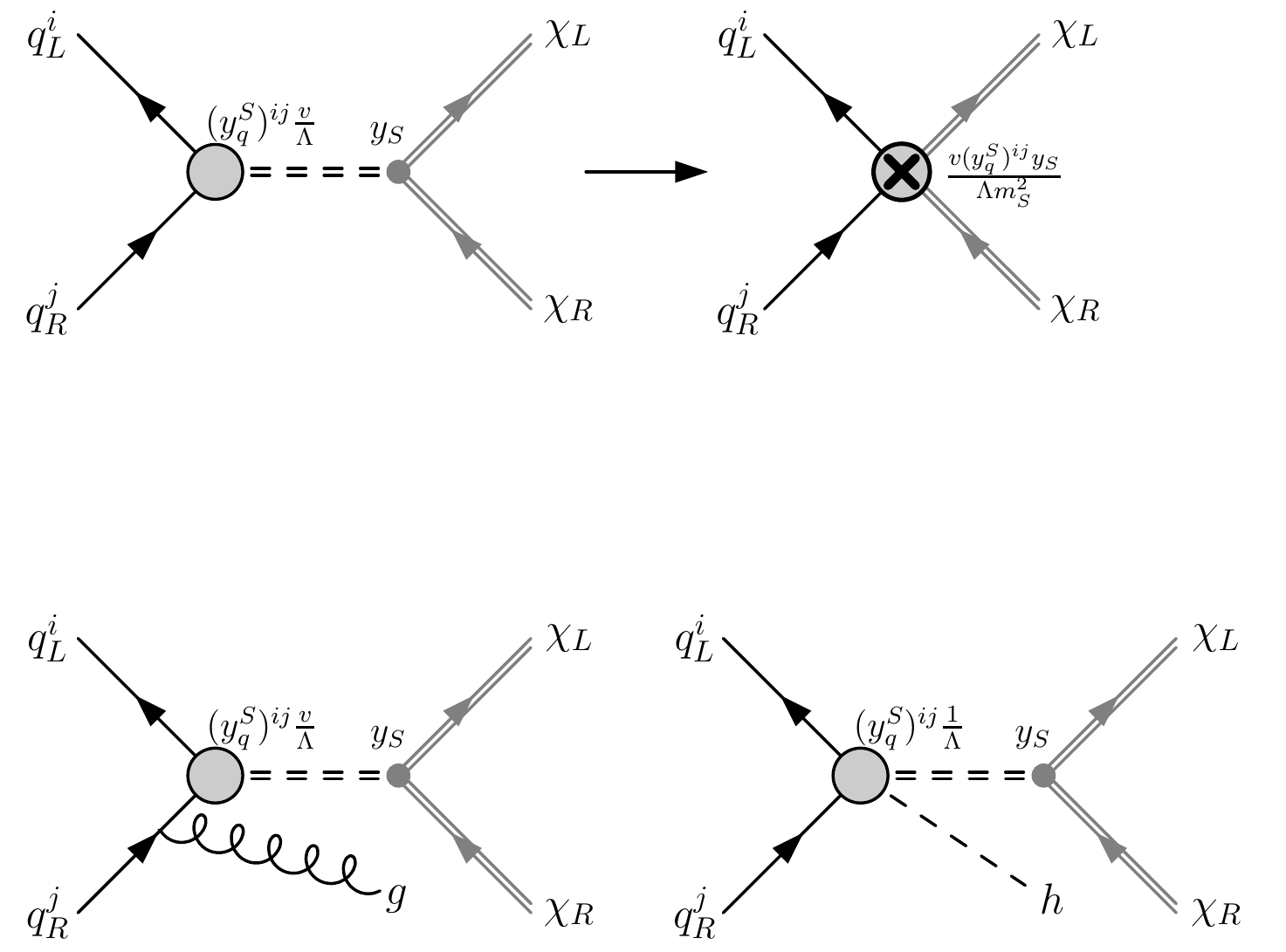} 
	\caption{\label{fig:diags} Relevant diagrams contributing to nuclear interaction with fermionic DM (first row)
	and corresponding DM observables at hadron colliders (monojet and Higgs$+\MET$, second row), turning
	on the interactions $\sim (y_q^S)^{ij}$ and $\sim y_S$. 
	The diagrams are similar for scalar and pseudo-scalar mediators, where for the latter
	case, the operator coefficients in Eq.~(\ref{eq:LEFT}) are to be replaced by the corresponding (tilded) 
	coefficients in Eq.~(\ref{eq:LEFT2}). See text for details.}
    \end{center}
\end{figure}

    The \eDMEFT\ Lagrangian, Eq.\eqref{eq:LEFT}, allows to describe phenomena relevant for collider searches 
    for DM as well as for direct (and indirect) detection experiments, as we will now explore in more detail.
    For example, for a non-negligible $(y_q^S)^{ij}/\Lambda$, the operators $S \bar{Q}_L^i H q_R^j$ can 
    mediate interactions of DM with a nucleus, 
    coupling the mediator to the DM via the 
    $S \bar \chi \chi$ interaction, see the upper panel
    of Fig.~\ref{fig:diags}, where the scalar mediator is depicted by double-dashed lines, while the fermionic DM
    is represented by faint double-lines. The same combination of operators induces, 
    on the other hand, DM signals at the LHC in two different, but correlated, incarnations. 
    Mono-jet and Higgs+missing transverse energy $(\MET)$ signatures are generated, radiating off a gluon
    from the (initial state) quarks and considering the physical Higgs within $H$ in the $S \bar{Q}_L^i H q_R^j$ operator, resp., 
    while the $S \to \bar \chi \chi$ transition 
    is responsible for the $\MET$, as shown in the lower panel of Fig.~\ref{fig:diags}.\footnote{
    At low (nuclear) energies, we can also 
    integrate out the mediator to arrive at a four-fermion interaction, as considered in the usual DM
    EFT~\cite{Beltran:2010ww,Bai:2010hh,Goodman:2010ku,Rajaraman:2011wf}. Now, however, the effective coefficient is fixed by Eq.~\eqref{eq:LEFT}.
    }  
   
    Note that, since the Yukawa-like couplings $\sim (y_q^S)^{ij}$ can feature flavour-changing 
    neutral currents (FCNCs), a flavour-protection mechanism like minimal flavour violation could be thought of,
    which would lead to a suppressed coupling to the light valence quarks and thus small effects at colliders
    at the tree level (see also Ref.~\cite{Goertz:2014qia}). In an agnostic approach, however, all couplings
    could be treated as free, and some tuning withing the Yukawa structure could be allowed, 
    considering only `direct' experimental constraints allowing, in principle, for considerable effects for valence quarks.
    Finally note that, in case light-quark contributions were suppressed, the operators could still 
    induce a coupling to gluons at the one-loop level, via heavy-quark triangle diagrams.

\begin{figure}[!t]
    \begin{center}
	\includegraphics[height=2.75in]{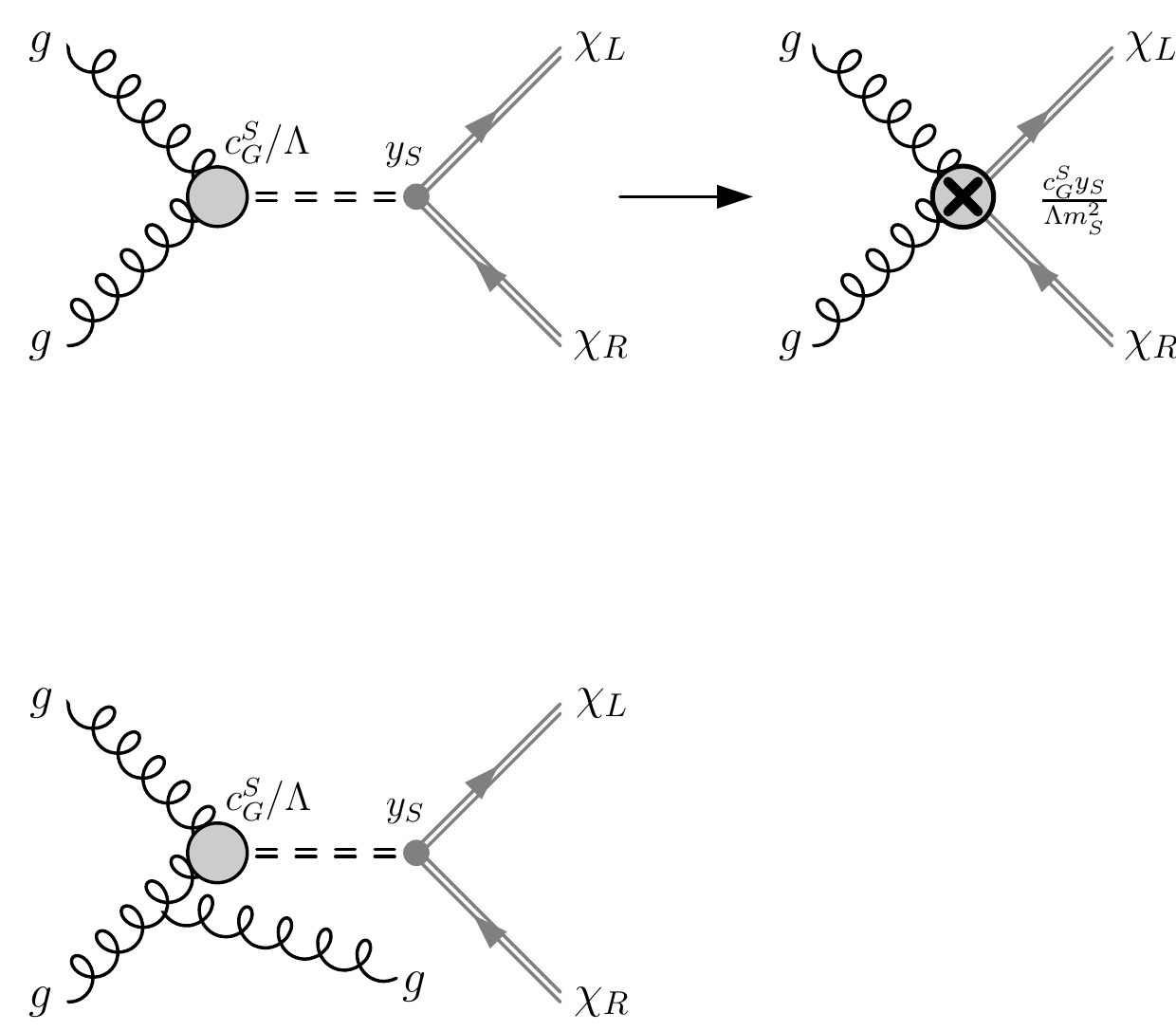} 
	\caption{\label{fig:diags2} Relevant diagrams contributing to nuclear interaction with fermionic DM (first row)
	and corresponding DM observables at hadron colliders (monojet, second row), turning
	on the interactions $\sim c_G^S$ and $\sim y_S$. 
	The diagrams are similar for pseudo-scalar mediators, employing the corresponding 
	tilded coefficients, as discussed before.}
    \end{center}
\end{figure}

    If other $D=5$ operators (as well as the portal coupling, $\lambda_{HS}^\prime$)  
    are set to zero for the moment, all three processes above scale in terms of effective coefficients
    as $y_S (y_q^S)^{ij}/\Lambda$, and we can explore the complementarity and combined information of 
    both types of experiments in one framework. We will examine this in more detail in the next section. 
    As an alternative option, we will  also consider the coupling of the mediator to the proton/nucleus via
    the ${\cal S} G^{a\mu\nu}G_{\mu\nu}^a$ operator, trading $(y_q^S)^{ij}$ for $c_G^S$,
    which allows for the production of DM in gluon fusion.
    The corresponding diagrams for the processes discussed above 
    are shown in Fig. \ref{fig:diags2}, where similar correlations can be explored (with the Higgs-associated 
    production now being absent at leading order). 
    In the same context, another interesting opportunity is to produce the mediator in weak-boson fusion (WBF),
    by turning on $c_W^S$ or $c_B^S$, as depicted by the diagrams in the last row
    of Fig. \ref{fig:diags3} and commented on further below.

\begin{figure}[!t]
    \begin{center}
	\includegraphics[height=2.6in]{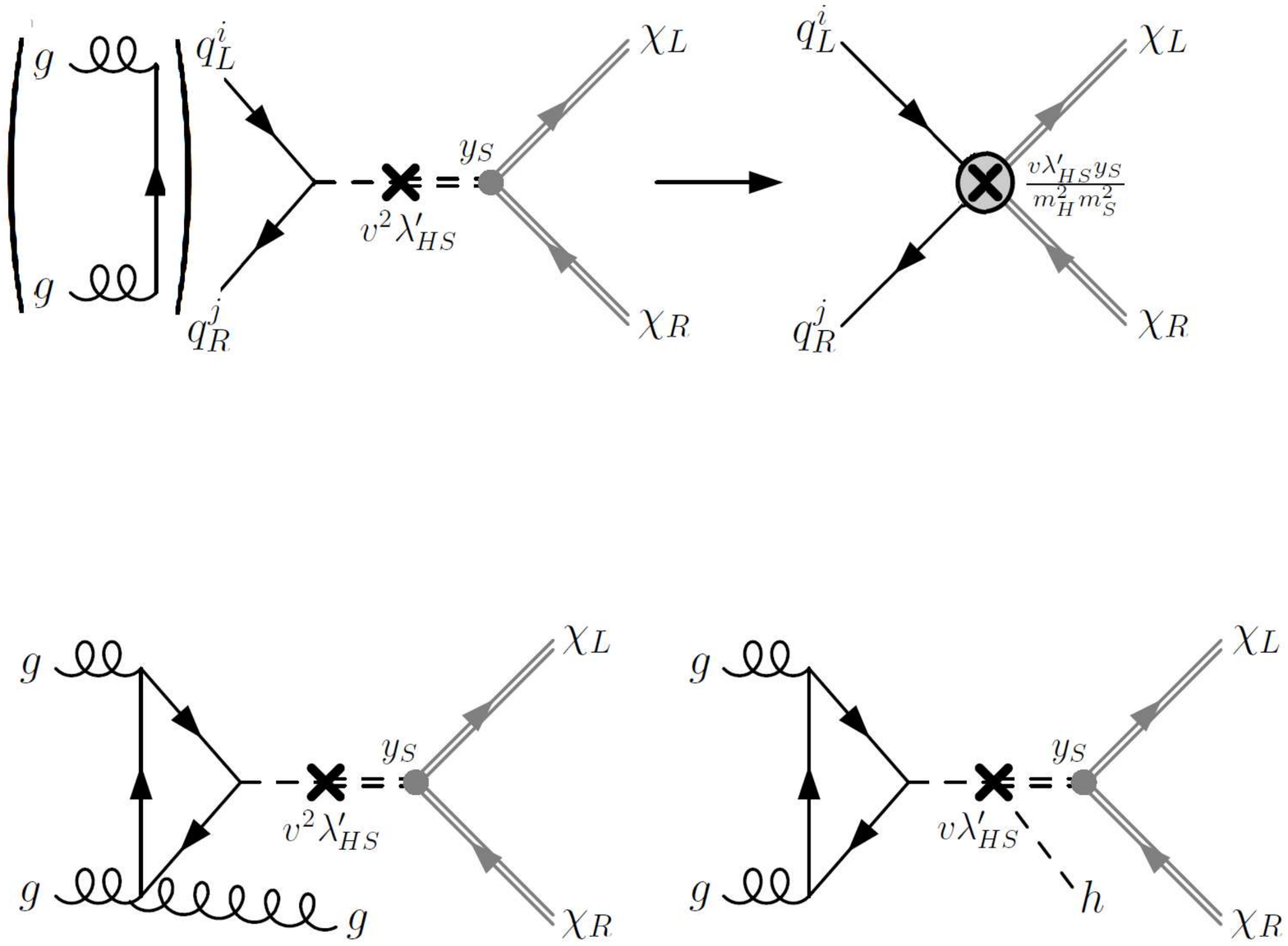} 
	\caption{\label{fig:diags4} Relevant diagrams contributing to nuclear interaction with fermionic DM (first row)
	    and corresponding DM observables at hadron colliders (monojet and Higgs$+\MET$, second row), turning
	    on the interactions $\sim \lambda_{HS}^\prime$ and $\sim y_S$. 
	    Note that the corresponding diagrams are not present for pseudo-scalar mediators, which only
	    appear in pairs.}
    \end{center}
\end{figure}

\begin{figure}[!t]
    \begin{center}
	\includegraphics[height=2.6in]{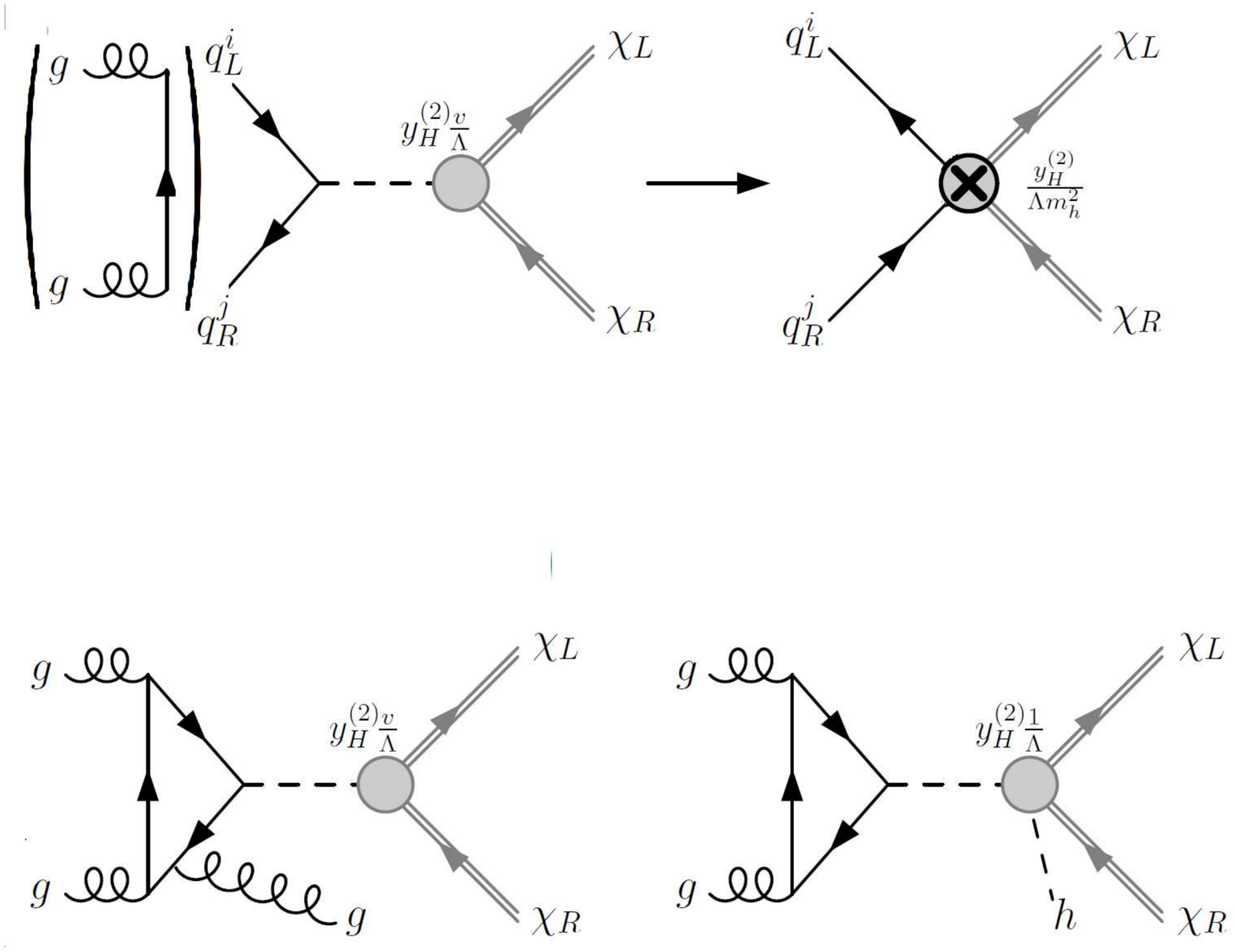} 
	\caption{\label{fig:diags5} Relevant diagrams contributing to nuclear interaction with fermionic DM (first row)
	    and corresponding DM observables at hadron colliders (monojet and Higgs$+\MET$, second row), turning
	    on the interaction $\sim y_H^{(2)}$.}
    \end{center}
\end{figure}

    On the other hand, the Higgs boson might play a crucial role in coupling the DM to quarks 
    and gluons. First, it can provide a portal to the mediator, via the couplings $\lambda_{HS},
    \lambda^\prime_{HS}$, connecting the SM to the dark sector.
    In particular if the second operator is present, the mediator can be produced in gluon-fusion
    Higgs production via mixing with the Higgs field 
    connecting then to the DM via $y_S$. 
    The corresponding diagrams are given in Fig.~\ref{fig:diags4},
    where again unavoidably the Higgs$+\MET$ channel is present, fixed by gauge invariance.
    Finally, the operators in Eq.~\eqref{eq:LEFT} also allow for interactions of DM with
    hadrons mediated {\it directly} by Higgs exchange, if the coefficient $y_H^{(2)}$
    is non-vanishing. Turning on this single coupling provides an instantaneous link 
    between the Higgs field and the DM via a contact interaction, inducing all
    processes discussed before, with the diagrams given in Fig.~\ref{fig:diags5}.
    Once more, the Higgs$+\MET$ channel is induced by gauge invariance.

    Let us conclude this discussion by emphasizing again that the Lagrangian of Eq.~\eqref{eq:LEFT}
    allows to consistently combine various processes and to include information from different kinds of sources.
    For example, DM might be produced by a combination of different mechanisms,
    e.g. via the mediator ${\cal S}$ (triggered by $(y_q^S)^{ij}$, $c_G^S$,
    or a portal), but also via direct Higgs exchange---due to the effective $y_H^{(2)}$, even without
    $H- \cal S$ mixing---where each contribution leads to characteristic correlations between LHC physics and
    direct (as well as indirect) detection experiments. We note that the direct Higgs couplings also enter
    in invisible Higgs decays constraining their size, while resonance searches are directly sensitive 
    to the properties of the mediator.
    The \eDMEFT\ allows to describe and combine all these 
    different phenomena in a 
    general (inclusive) and consistent way including resonance searches for the mediator particle,
    which would not be possible in a simple DM EFT or a simplified-model approach. 
    Yet, it is simple 
    enough to keep predictivity and to be straightforwardly implemented into tools for automated
    event generation and implementation of constraints.
    Finally, matching UV complete models to the \eDMEFT\ will also allow
    to interpret experimental results obtained in the latter framework in terms of such 
    explicit models, without the need to repeat the analysis for numerous different setups.

    Before exploring in more detail the LHC and direct-detection phenomenology 
    in the \eDMEFT\ approach,
    we will now turn to the remaining scenarios of DM coupled to the SM
    via a potentially light mediator, which can be analyzed in a similar way.
    First, we will consider the case of a CP-odd scalar, $\tilde {\cal S}$,
    which has several interesting features.
    On the one hand, direct detection bounds with a CP-odd scalar mediator are much 
    weaker due to momentum suppression of the cross section~\cite{Boehm:2014hva,Arina:2014yna}.
    Moreover, besides the $D=5$ pure scalar interactions, the portal with a single 
    mediator vanishes, which automatically avoids the mixing between the scalars.
    The 
    Lagrangian now becomes
    \begin{eqnarray}
    \label{eq:LEFT2}
    \mathcal{L}_{\rm eff}^{\tilde{\cal S} \chi}& = & {\cal L}_{\rm SM} +
    \frac{1}{2}\partial_\mu \tilde{\cal S} \partial^\mu \tilde{\cal S} 
    - \frac 1 2 \mu_{\tilde{S}}^2 \tilde{\cal S}^2 + \bar \chi i \dslash\chi- m_\chi \bar \chi \chi \nonumber \\
    &-& \frac{\lambda_{\tilde S}}{4} \tilde{\cal S}^4 - \lambda_{H\tilde{S}} |H|^2 \tilde{\cal S}^2 \ \,
    - y_{\tilde S}\,  \tilde{\cal S } i \bar \chi_L \chi_R +\mathrm{h.c.} 
    \nonumber \\[2.5mm]
    &-& \frac{\tilde{\cal S}}{\Lambda} \left[(y_d^{\tilde S})^{ij} i \bar{Q}_L^i H d_R^j + (y_u^{\tilde S})^{ij} i \bar{Q}_L^i\tilde{H}u_R^j \right.\nonumber\\
    && \quad \ + \left.(y_\ell^{\tilde S})^{ij} i\bar{L}_L^i H \ell_R^j +\mathrm{h.c.}\right]\\[1mm]
    &-&\frac{y_{\tilde S}^{(2)} \tilde{\cal S}^2 + y_H^{(2)} H^\dagger H}{\Lambda}\bar{\chi}_L \chi_R +\mathrm{h.c.}
    \nonumber\\[1mm]
    &-&\frac{\tilde{\cal S}}{\Lambda}\frac{1}{16\pi^2}\left[g^{\prime 2} c_B^{\tilde S} B_{\mu\nu} \tilde B^{\mu\nu}+ g^2 c_W^{\tilde S} 
    W^{I\mu\nu} \tilde W_{\mu\nu}^I\right.\nonumber\\
    && \qquad \qquad +\left.g_s^2 c_G^{\tilde S} G^{a\mu\nu} \tilde G_{\mu\nu}^a\right]\,.\nonumber
    \end{eqnarray}

    Here, the contact interactions with gauge bosons feature dual field strength tensors,
    while all terms can be interpreted analogously as in the CP-even scalar case.
    In particular, the very same discussion as before on generic correlations between different
    observables can be performed. Note, however, that due to the vanishing
    single-mediator portal, no production via the Higgs is possible at the level of $D \leq 4$
    interactions, which makes the (extended) EFT terms $\sim y_q^{\tilde S}, c_{G\!,B\!,W}, 
    y_H^{(2)}$ even more interesting. We will again assume no new sources of CP violation, and the coefficients in Lagrangian in 
    Eq.~\eqref{eq:LEFT2} are real.

    \subsection{Scalar DM with a scalar 
    mediator}

    We finally move to the case of (singlet) scalar DM, denoted by ${\cal D} = \chi_s$ still considering a scalar mediator,
    ${\cal M} = {\cal S}$. The Lagrangian for this setup, at the $D=5$ level, reads     
    \begin{equation}
    \label{eq:LEFT3}
    \begin{split}
    \mathcal{L}_{\rm eff}^{{\cal S} \chi_s} = & {\cal L}_{\rm SM} +
    \ \frac{1}{2}\partial_\mu {\cal S} \partial^\mu {\cal S} + \frac{1}{2}\partial_\mu  \chi_s \partial^\mu \chi_s- \lambda_{S1}^{\prime}v^3{\cal S}\\
    &-\frac 1 2 \mu_S^2 {\cal S}^2 - \! \frac 1 2 m_{\chi_s}^2 \chi_s^2 
    -\frac{\lambda^{\prime}_S}{2\sqrt{2}} v{\cal S}^3- \frac{\lambda_{S}}{4} {\cal{S}}^4\\
    & - \frac{\lambda_{\chi_s}}{4} \chi_s^4
    - \lambda^\prime_{HS} v |H|^2 {\cal{S}} 
	- \lambda_{HS} |H|^2 {\cal{S}}^2  \\
    &- \frac{\lambda^\prime_{S \chi_s}}{2 \sqrt 2} v\, {\cal S} \chi_s^2 - \lambda_{S\chi_s} {\cal S}^2 \chi_s^2   
	- \lambda_{H\chi_s} |H|^2 \chi_s^2  \\[2.5mm]
    &- \frac{\cal{S}}{\Lambda} \left[ c_{\lambda S} {\cal{S}}^4 
    +c_{HS} |H|^2 {\cal{S}}^2 
    +c_{\lambda H} |H|^4 \right. \\
    & \quad+ \left.
     c_{S \chi_s} {\cal S}^2 \chi_s^2  + c_{\lambda \chi_s} \chi_s^4 
    +c_{H\chi_s} |H|^2 \chi_s^2 
    \right] \\[1mm]
    &- \frac{\cal{S}}{\Lambda} \left[(y_d^S)^{ij} \bar{Q}_L^i H d_R^j + (y_u^S)^{ij}\bar{Q}_L^i\tilde{H}u_R^j \right.\\
    & \quad \ + \left.(y_\ell^S)^{ij} \bar{L}_L^i H \ell_R^j +\mathrm{h.c.}\right]\\[1mm]
    &-\frac{\cal{S}}{\Lambda}\frac{1}{16\pi^2}\left[g^{\prime 2} c_B^S B_{\mu\nu} B^{\mu\nu}+ g^2 c_W^S W^{I\mu\nu} W_{\mu\nu}^I\right.\\
    & \qquad \qquad +\left.g_s^2 c_G^S G^{a\mu\nu}G_{\mu\nu}^a\right]\,. 
    \end{split}
    \end{equation}
    We assume again that 
    the mediator, $\cal S$, does not develop a vev, 
    while a $Z_2$ symmetry assures stability of~$\chi_s$.
    
    For pseudo-scalar, ${\cal M}=\tilde{{\cal S}}$, the terms in the last two square brackets are replaced
    similarly as before (Eq.~\eqref{eq:LEFT}~$\to$~Eq.~\eqref{eq:LEFT2}). However, all further
    contributions with an odd number of ${\cal S}$ are absent, and further dynamics would be required for $s$-channel 
    mediation between the DM and SM fields, and therefore, we do not consider this scenario here.

    Crucial (new) terms in Eq.~\eqref{eq:LEFT3} 
    are the portal-like interactions
    connecting the DM to the mediator,  $\lambda_{S \chi_s}$, 
    $\lambda^\prime_{S \chi_s}$ 
    (and the corresponding $D=5$ operators, containing odd powers of ${\cal S}$ and even powers of $\chi_s$, 
    i.e., the terms $\sim c_{S \chi_s},c_{\lambda \chi_s}, c_{H \chi_s}$), replacing $y_S/y_{\tilde S}$
    of the fermionic DM case.
    Furthermore, there is a new direct $D=4$ portal to the DM via the Higgs field, given by $\lambda_{H \chi_s}$, 
    instead of the corresponding term $\sim y_H^{(2)}$ for fermionic DM,
    as well as a quartic DM self-interaction $\sim \lambda_{\chi_s}$. 
    We will 
    discuss the corresponding changes in the processes described above for fermionic dark matter, including the
    respective diagrams, in the next section.

\section{Complementary constraints within one 
framework: $\text{eDM{\scriptsize EFT}}$}
\label{sec:pheno}

We will now explore phenomenological aspects of all
the models described above, making use of the correlations predicted in our \eDMEFT\ approach, 
as anticipated before.   In particular, we will address the question of how many events in DM searches at the LHC can be expected,
given limits from direct detection experiments, without restricting to an explicit NP model. The presence
of the mediator in our EFT will ensure the validity of the analysis for collider searches.
We will finally discuss the inclusion of further processes like
loop-induced DM production via $D=5$ operators,
invisible Higgs decays, Higgs pair production, as well as resonance searches in the same framework.

As main observables, we consider the direct-detection cross section for the scattering of DM
off a nucleus, $N$, 
\begin{equation}
    \sigma_N \equiv \sigma(N {\cal D} \to N {\cal D}) \,,
\end{equation}
and the cross section for a monojet signal at the 13~TeV 
LHC 
\begin{equation}
    \sigma_j  \equiv \sigma(p p \to j + \MET) \,.
\end{equation}
In addition, in the cases where the mediator is produced via the new $D=5$ operators coupling it to  quarks, we expect also the 
cross section for producing the Higgs in association with missing transverse energy to be comparable to the monojet cross section and consider also
\begin{equation}
    \sigma_{h\!+\!\MET} \equiv \sigma(p p \to h + \MET)  \,.
\end{equation}
We can now calculate how many events in LHC DM searches (monojet and Higgs$+\MET$) are
possible, given the limits from direct detection, in the framework of different {\eDMEFT}s  
discussed in the last section, each representing a large class of NP models.
The LHC results could thus potentially rule out certain DM incarnations very generally or support/constrain them in synergy with 
direct-detection experiments.
In that context, note that the currently most stringent direct detection constraint corresponds to $N = {\rm Xe}$
and is given by XENON1T~\cite{Aprile:2017iyp}.

Going through the different scenarios, we will address this question,
always turning on a set of (one or two) NP couplings that allow
for a different production mechanism of the DM or different DM--nucleus interaction,
considering quark- and gluon-induced production, production through the Higgs-mediator portal,
and production trough direct Higgs--DM interactions.

The requirement to generate the correct relic abundance implies further constraints
on the same operators that are entering DM--SM scattering and thus direct detection. 
On the other hand, the limits from direct detection for a subdomimant DM component are 
weaker, since the direct detection experiments assume a one-component DM with the observed
relic abundance. Therefore, in order to correctly compare with the limits, the produced DM abundance
must always be estimated. 

Another source of constraints arises from DM annihilations potentially producing an excess of e.g. gamma rays over the galactic background. 
The most stringent current limits come from the Fermi-LAT satellite experimemt~\cite{Ackermann:2015zua} constraining the  canonical 
thermal cross section up to $\sim 100$~ GeV DM masses. We focus here on scenarios with heavier DM candidate and leave 
further analysis of indirect observables to future work.

    \subsection{Quark-induced production}

\begin{figure}[!t]
	    \begin{center}
		\includegraphics[height=2.6in]{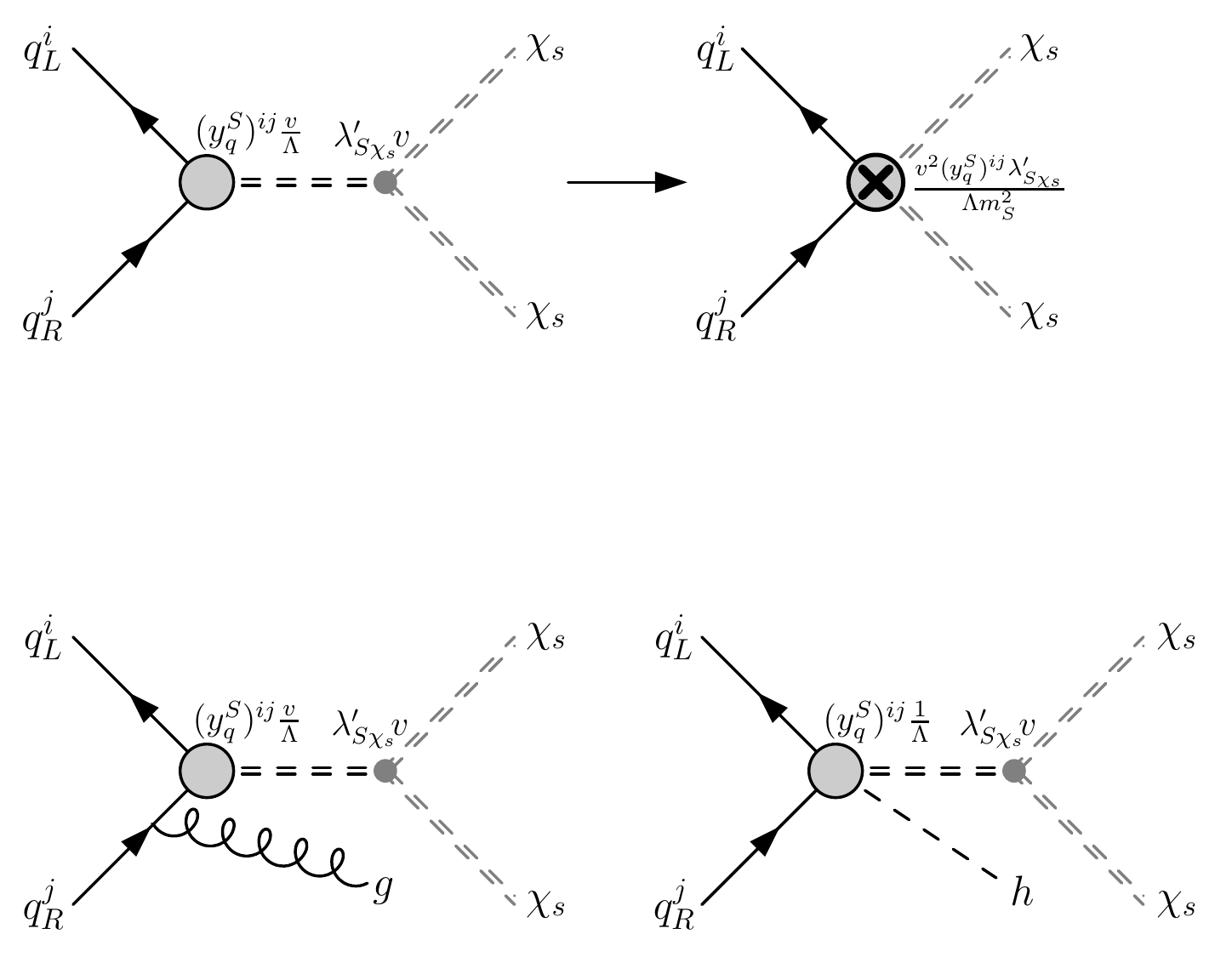} 
		\caption{\label{fig:diagsS1} Relevant diagrams contributing to nuclear interaction with scalar DM (first row)
		     and corresponding DM observables at hadron colliders (monojet and Higgs$+\MET$, second row), turning
		     on the interactions $\sim (y_q^S)^{ij}$ and $\sim \lambda_{S \chi_s}^\prime$.}
	    \end{center}
	\end{figure}

    We start by considering the production of the mediator, ${\cal M}=\cal S$, 
    in $q\bar q$
    annihilation and subsequent decay to DM (as well as the crossed process
    leading to ${\cal D}-N$ scattering), see Figs~\ref{fig:diags} and \ref{fig:diagsS1}.

	\subsubsection{Fermionic DM}

	For fermionic DM, ${\cal D}=\chi$, the relevant couplings 
	for a CP-even mediator, $\cal S$, are 
	$(y_q^S)^{ij}$ and $y_S$, see Eq. \eqref{eq:LEFT}.
	The former allows the production
	of ${\cal S}$ via a gauge-invariant coupling to SM quarks
	and the latter its decays to DM. The corresponding Feynman diagram
	is given in the upper left corner of Fig.~\ref{fig:diags}.
	At low energies relevant for direct-detection experiments,
	the mediator can be integrated out leading to the
	diagram in the upper right corner, which governs 
	${\cal D}-N$ scattering at low momenta. For 
	simplicity, we only consider $(y_u^S)^{11}$,
	although the analysis can easily be extended to include all quark flavours.

	For a CP-odd mediator, ${\cal \tilde S}$, the couplings above are replaced by
	the corresponding tilded coefficients in Eq.~\eqref{eq:LEFT2}. However, 
	in this case, the tree-level interactions with nuclei are momentum suppressed, and 
	this scenario is out of the reach of current direct detection experiments (and thus LHC cross sections
	are basically unconstrained).  
	Nevertheless, as discussed in Ref.~\cite{Arcadi:2017kky}, the future experiments will start probing
	this scenario as well. Scenarios with CP-odd mediator have also been considered recently in e.g. 
	Refs~\cite{Mambrini:2015wyu,Arcadi:2017wqi,Bauer:2017fsw}.
	Here, we concentrate on a CP-even mediator, and leave the phenomenology
	of CP-odd mediators, utilizing the full strength of the \eDMEFT\ approach, for future work.
	
	The cross section for the DM scattering off nuclei, in terms of \eDMEFT\ couplings, reads~\cite{Alanne:2014bra}
	\begin{equation}
	    \label{eq:}
	    \sigma_N=\frac{y_S^2 [(y_u^S)^{11})]^2 (f^u_N)^2 m_N^2\mu_N^2 v^2}{2\pi\Lambda^2 m_S^4 m_u^2},
	\end{equation}
	where $f^u_N$ is the form factor defined by ${\langle N|m_q \bar{q}q|N\rangle\equiv m_N f^q_N}$,  
	\begin{equation}
	    \label{eq:}
		\mu_N\equiv \frac{m_{\chi} m_N}{m_{\chi}+m_N}   
	\end{equation}
	is the reduced mass of the DM-nucleon system, and $m_N=(m_p+m_n)/2$ is the average nucleon mass.

	We calculate the running of the matrix elements $\langle N|y^S_{q}\bar{q}q|N\rangle$ 
	from the EW scale (where we define our couplings) down to direct-detection energies and the 
	corresponding threshold effects from integrating out the heavy quark flavours
	following Ref.~\cite{Hill:2014yxa} (see also\cite{Bishara:2017pfq,Bishara:2017nnn}). 
	For the case with only $(y_u^S)^{11}$ non-zero, this effect is trivial. 
	
	We fix $y_S$ such that we obtain the fraction ${f_{\mathrm{rel}}\equiv \Omega_{\chi}/\Omega_{\mathrm{DM}}}$ 
	of the total dark matter abundance. In the direct-detection limits, a single component DM with $f_{\mathrm{rel}}=1$
	is assumed, and thus we require $f_{\mathrm{rel}}\sigma_N\leq \sigma_{\mathrm{X1T}}$ when we compare with the limits 
	for XENON1T. 

	If $m_{\chi}<m_S$, the DM annihilation cross section is dominated by the $s$-channel process to SM quarks via the $D=5$ operator, and 
	the cross section is thus proportional to $y_S^2 [(y_u^S)^{11}]^2$. This is the same combination appearing in the direct-detection
	cross section and values avoiding overabundance are disfavoured by experiments (which does not necessarily hold if the DM couples to
	heavy quarks). 
	Therefore, we concentrate here on the mass range 
	$m_{\chi}>m_S$, where the dominant annihilation channel is $\chi\chi\rightarrow SS$. The annihilation cross section 
	can thus be estimated (neglecting subleading $m_S$ contributions, which are considered in the numerical results) by~\cite{Arcadi:2017kky} 
	\begin{equation}
	    \begin{split}
		\label{eq:frelfDM}
		&\langle\sigma \vel \rangle(\chi\chi\rightarrow SS)
		\approx 2.0\times 10^{-26}\, \mathrm{cm}^3 \mathrm{s}^{-1}y_S^4\left(\frac{1\, \mathrm{TeV} }{m_{\chi}}\right)^2,   
	    \end{split}
	\end{equation}
	where $\vel^2 \sim 0.23$. 
	We fix $y_S$ as a function of $f_{\mathrm{rel}}$ by comparing this to the standard thermal cross section 
	$\langle\sigma \vel\rangle_0=3\cdot 10^{-26}\ \mathrm{cm}^3 \mathrm{s}^{-1}$ and using $\Omega h^2\propto \langle\sigma \vel\rangle^{-1}$
	to scale this with $f_{\mathrm{rel}}$.
	The latest limit~\cite{Aprile:2017iyp} then leads to the bound (see Fig.~\ref{fig:yuqlimits})
	\begin{equation}
	    \label{eq:limqf}
	    \begin{split}
		\frac{|(y_u^S)^{11}|}{\Lambda} \lesssim  2.9\times 10^{-3}f_{\mathrm{rel}}^{-1/4}
		    \left(\frac{m_S}{1\ \mathrm{TeV}}\right)^2\, \mathrm{TeV}^{-1}.
	    \end{split}
	\end{equation}
		
	Attaching on the other hand a gluon to the initial state
	quarks or emitting a Higgs boson---possible directly
	via the contact interaction $\sim (y_q^S)^{ij}$---leads 
	to final states considered in LHC searches for
	DM,  i.e. monojet and Higgs$+\MET$ signatures.
	The corresponding diagrams are given in the lower panel
	of Fig.~\ref{fig:diags}.
	
	We calculate the maximal cross sections for these processes at the LHC at 13~TeV center-of-mass energy 
	using MadGraph~\cite{Alwall:2014hca}
	and employing the direct-detection limit, Eq.~\eqref{eq:limqf}, for two benchmark scenarios: 
	$(m_S=400\, \mathrm{GeV}, m_{\chi_s}=500\, \mathrm{GeV})$ and 
	$(m_S=500\, \mathrm{GeV}, m_{\chi_s}=1\, \mathrm{TeV})$.   
	We arrive at (fixing for simplicity $f_{\mathrm{rel}}=1$) 
	\begin{equation}
	    \begin{split}
		&\sigma_j|_{m_{\chi}=500\,\mathrm{GeV}} \lesssim 3.0\cdot 10^{-7}\, \mathrm{fb},\\ 
		&\sigma_j|_{m_{\chi}=1\,\mathrm{TeV}} \lesssim 3.5\cdot 10^{-8}\, \mathrm{fb}, 
	    \end{split}
	\end{equation}
	and
	\begin{equation}
	    \begin{split}
		&\sigma_{h+\MET}|_{m_{\chi}=500\,\mathrm{GeV}} \lesssim 2.0\cdot 10^{-8}\, \mathrm{fb},\\ 
		&\sigma_{h+\MET}|_{m_{\chi}=1\,\mathrm{TeV}} \lesssim 3.4\cdot 10^{-8}\, \mathrm{fb}. 
	    \end{split}
	\end{equation}
	These cross sections are tiny, but serve here as reference values for this utterly simplified scenario with only two
	addtional couplings turned on. For example, already allowing for heavy quark flavors to couple to the mediator, interpolating 
	between the light quark case and the gluon case discussed in the next section, is expected
	to increase the cross section significantly.

	\subsubsection{Scalar DM}

	For scalar DM, ${\cal D}=\chi_s$, the corresponding couplings are
	still $(y_q^S)^{ij}$ for the $q\bar q$ production of the mediator,
	while the decay to DM is now induced by the portal term
	$\lambda_{S \chi_s}^\prime$.
	The Feynman diagrams for the processes
	discussed above are analogously presented
	in Fig.~\ref{fig:diagsS1}, where $\chi_s$ is represented
	by faint dashed double-lines. Note that for a CP-odd mediator 
	the above portal linear in $\tilde S$ is not present. 

	\begin{figure}
	    \begin{center}
		\includegraphics[width=0.45\textwidth]{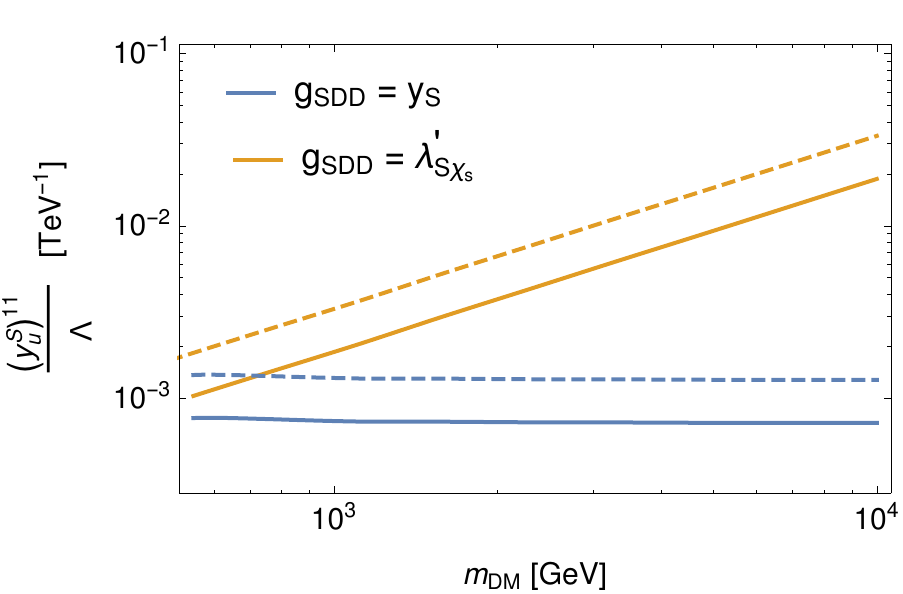}
	    \end{center}
	    \caption{Limits from the direct-detection experiments for the coupling $(y_u^S)^{11}/\Lambda$ as a function of the DM mass.  
	    Blue (yellow) lines correspond to fermionic (scalar) DM with CP-even mediator, and $g_{S\chi\chi}$ denotes 
	    the corresponding mediator--DM coupling. 
	    In the plot, we have fixed $m_S=500$~GeV, and the 
	    solid (dashed) lines correspond to $f_{\mathrm{rel}}=1$ ($f_{\mathrm{rel}}=0.1$).}
	    \label{fig:yuqlimits}
	\end{figure}

	The scattering cross section off nuclei now becomes~\cite{Alanne:2014bra}
	\begin{equation}
	    \label{eq:}
	    \sigma_N=\frac{(\lambda_{S\chi_s}^{\prime})^2 (y^S_{11})^2 (f^u_N)^2 m_N^2\mu_N^2 v^4}{16\pi\Lambda^2 m_S^4 m_u^2 m_{\chi_s}^2}.
	\end{equation}
	
	Again, to estimate the relic density, we concentrate on $m_{\chi_s}>m_S$, whence the 
	annihilation cross section yields~\cite{Arcadi:2017kky}
	\begin{equation}
	    \begin{split}
		\label{eq:frelSDM}
		\langle\sigma \vel \rangle(\chi_s\chi_s\rightarrow SS)
		\approx& 5.8\times 10^{-26}\, \mathrm{cm}^3 \mathrm{s}^{-1}\\
		&\cdot\left(\frac{\lambda_{S\chi_s}^\prime v}{2\sqrt{2}m_S}\right)^4
		    \left(\frac{1\, \mathrm{TeV} }{m_{\chi_s}}\right)^2.  
	    \end{split}
	\end{equation}
	Trading $\lambda_{S\chi_s}^{\prime}$ for $f_{\mathrm{rel}}$ similarly as above, 
	we show  the bounds from direct detection again in Fig.~\ref{fig:yuqlimits}.
	For the benchmark scenarios $(m_S=400\, \mathrm{GeV}, m_{\chi_s}=500\, \mathrm{GeV})$ and 
	$(m_S=500\, \mathrm{GeV}, m_{\chi_s}=1\, \mathrm{TeV})$,  we obtain limits
	\begin{equation}
	    \label{eq:limq}
	    \begin{split}
		&\frac{|(y_u^S)^{11}|}{\Lambda}|_{m_{\chi_s}=500\, \mathrm{GeV}}
		    \lesssim  7.5\times 10^{-4}f_{\mathrm{rel}}^{-1/4}\, \mathrm{TeV}^{-1},\\
		&\frac{|(y_u^S)^{11}|}{\Lambda}|_{m_{\chi_s}=1\, \mathrm{TeV}}
		    \lesssim  1.9\times 10^{-3}f_{\mathrm{rel}}^{-1/4}\, \mathrm{TeV}^{-1}.
	    \end{split}
	\end{equation}
		
	Using these and fixing $f_{\mathrm{rel}}=1$, the maximal values of the LHC cross sections become
	\begin{equation}
	    \begin{split}
	    &\sigma_j|_{m_{\chi_s}=500\,\mathrm{GeV}} \lesssim 1.2\cdot 10^{-7}\, \mathrm{fb},\\ 
	    &\sigma_j|_{m_{\chi_s}=1\,\mathrm{TeV}} \lesssim 6.8\cdot 10^{-9}\, \mathrm{fb}, 
	    \end{split}
	\end{equation}
	and
	\begin{equation}
	    \begin{split}
		&\sigma_{h+\MET}|_{m_{\chi_s}=500\,\mathrm{GeV}} \lesssim 1.6\cdot 10^{-8}\, \mathrm{fb},\\ 
		&\sigma_{h+\MET}|_{m_{\chi_s}=1\,\mathrm{TeV}} \lesssim 8.1\cdot 10^{-10}\, \mathrm{fb}. 
	    \end{split}
	\end{equation}
	which again are tiny as expected for this simplified scenario.

	\subsection{Gluon-fusion production}

\begin{figure}[!t]
	    \begin{center}
		\includegraphics[height=2.6in]{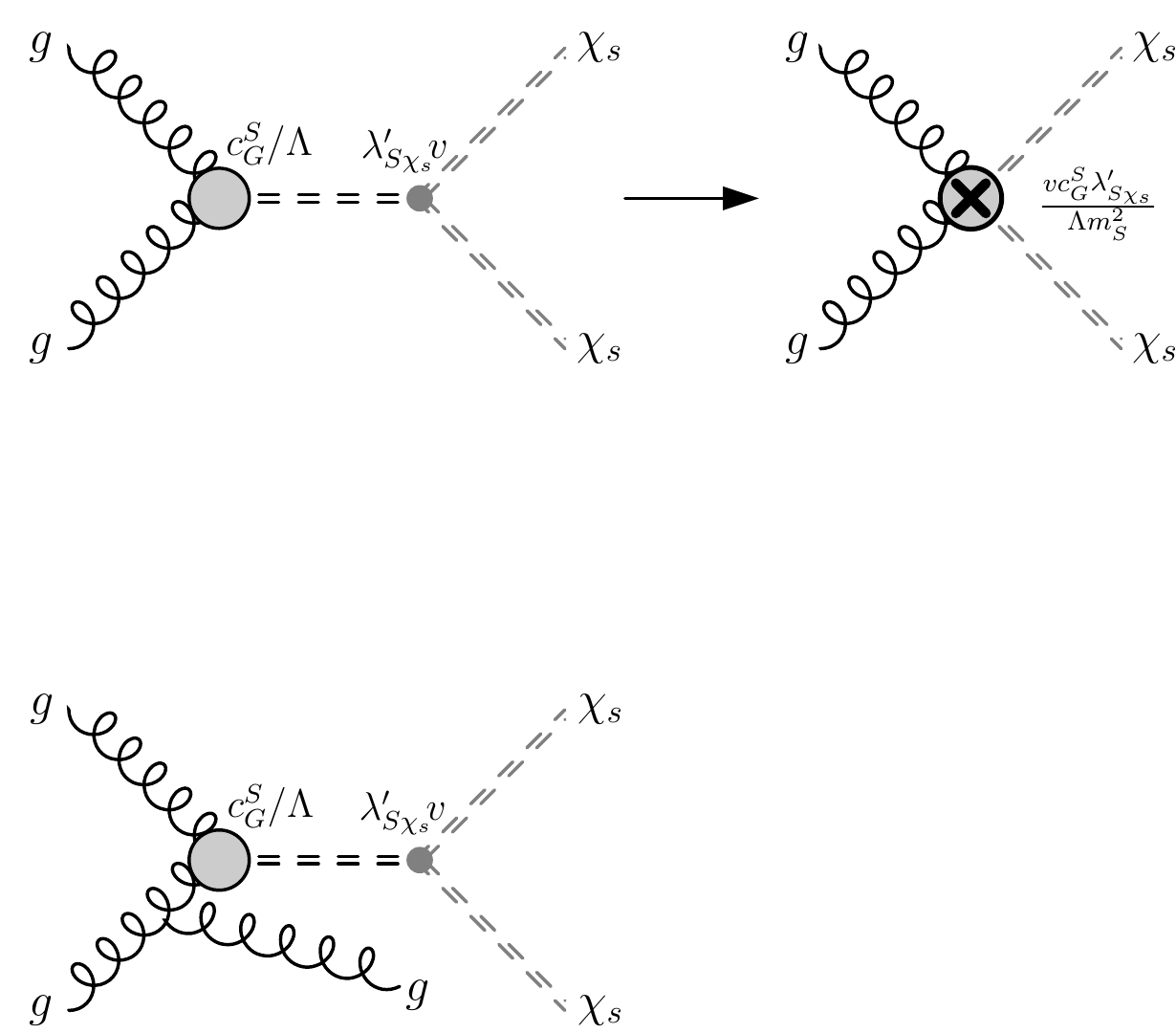} 
		\caption{\label{fig:diagsS2} Relevant diagrams contributing to nuclear interaction with scalar DM (first row)
		     and corresponding DM observables at hadron colliders (monojet, second row), turning
		     on the interactions $\sim c_G^S$ and $\sim \lambda_{S \chi_s}^\prime$.
		     }
	    \end{center}
	\end{figure}

	We now turn to the case of coupling the mediator to a $gg$ state,
	allowing its production in gluon-gluon fusion,
	which is complementary to the case of external $q \bar q$ states.
	Since the coupling to the DM is still assumed to be induced
	by $y_S$ 
	and $\lambda_{S \chi_s}^\prime$,
	for the cases of fermionic DM $\chi$ with a CP-even 
	mediator
	$S$ 
	and scalar DM $\chi_s$, respectively,
	we only replace ${(y_u^S)^{11}}$ by $c_G^S$ to couple the 
	mediator to gluons.
	The relevant diagrams are given in Figs~\ref{fig:diags2} and~\ref{fig:diagsS2}.
	Note that, due to the absence of the Higgs contact interaction in the NP
	sector, the Higgs$+\MET$ channel is significantly suppressed with respect
	to the monojet signature, requiring a $GG h$ interaction,
	induced in the SM via a top loop, and we do not consider that here.

	\subsubsection{Fermionic DM}

	Starting with the ${\cal S}GG$ interaction at the EW scale will now induce the ${\cal S}\bar{q}q$ couplings
	at the nuclear energy scale~\cite{Hill:2014yxa}. Therefore, the cross section for scattering off nuclei now becomes
	\begin{equation}
	    \label{eq:}
	    \sigma_N=\frac{y_S^2 (c^S_{G})^2  m_N^2\mu_N^2}{\pi \Lambda^2 m_S^4}
		\left(\sum_{q=u,d,s}(c_G^q f^q_N)+\frac{2}{9}c_G^g f^g_N\right)^2,
	\end{equation}
	where  $c_G^q$ and $c_G^g$ account for the running and threshold effects down to the  nuclear-energy scale, 
	the gluonic form factor is defined as ${\langle N|-\frac{g_s^2}{16\pi^2}\,G_{\mu\nu} G^{\mu\nu} |N\rangle\equiv \frac{2}{9} m_N f^g_N}$, and 
	we again employ the results of Ref.~\cite{Hill:2014yxa}.

\begin{figure}
	    \begin{center}
		\includegraphics[width=0.45\textwidth]{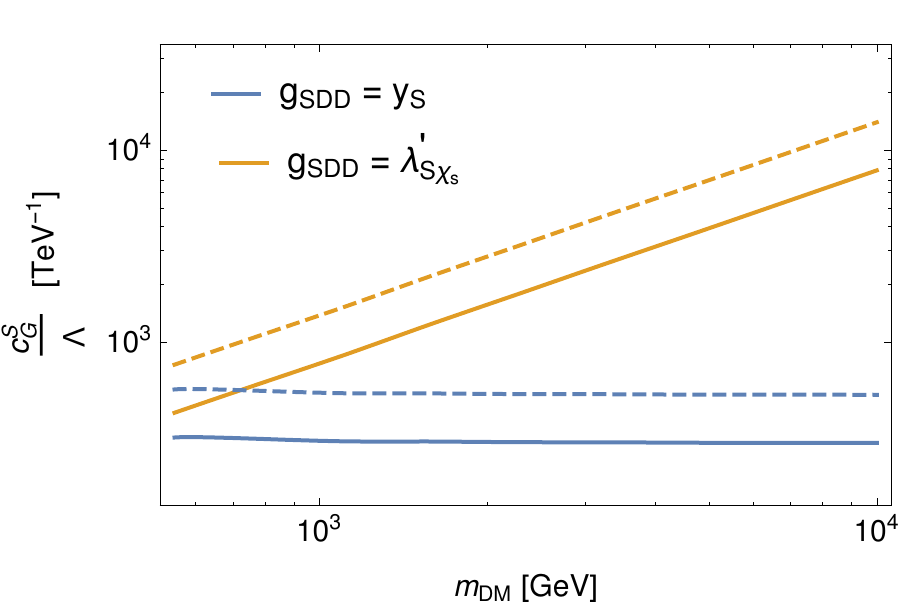}
	    \end{center}
	    \caption{Limits from the direct-detection experiments for the coupling $c_G^S/\Lambda$ as a function of the DM mass.  
	    Blue (yellow) lines correspond to fermionic (scalar) DM with CP-even mediator, and $g_{S\chi\chi}$ denotes the corresponding
	    mediator--DM coupling. In the plot, we have fixed $m_S=500$~GeV, and the 
	    solid (dashed) lines correspond to $f_{\mathrm{rel}}=1$ ($f_{\mathrm{rel}}=0.1$). 
	    }
	    \label{fig:cGbounds}
	\end{figure}

	Comparing now with XENON1T results, leads to the bound	
	\begin{equation}
	    \label{eq:limq}
	    \begin{split}
		\frac{c_G^S}{\Lambda} \lesssim  1.3\times 10^{3}f_{\mathrm{rel}}^{-1/4}
		    \left(\frac{m_S}{1\ \mathrm{TeV}}\right)^2\, \mathrm{TeV}^{-1}\,,
	    \end{split}
	\end{equation} shown in Fig.~\ref{fig:cGbounds}. 

	The monojet cross section
	is thus constrained for the benchmark scenarios $(m_S=400\, \mathrm{GeV}, m_{\chi}=500\, \mathrm{GeV})$ and 
	$(m_S=500\, \mathrm{GeV}, m_{\chi}=1\, \mathrm{TeV})$ and $f_{\mathrm{rel}}=1$ to
	\begin{equation}
	    \begin{split}
	    &\sigma_j|_{m_{\chi}=500\,\mathrm{GeV}} \lesssim 1.9\cdot 10^3\, \mathrm{fb},\\ 
	    &\sigma_j|_{m_{\chi}=1\,\mathrm{TeV}} \lesssim 250\, \mathrm{fb}. 
	    \end{split}
	\end{equation}
	These values are significantly higher than in the quark-induced production, connected to the relative growth of the gluon 
	parton distribution functions with respect to the up-quark one from nuclear to collider energies 
	and the negative interference between the quark and gluon form factors at nuclear scales. Therefore this scenario 
	provides an interesting possibility to partially evade the strong direct-detection limits.

	\subsubsection{Scalar DM}

	For scalar DM (with a CP-even mediator), we obtain
	\begin{equation}
	    \label{eq:}
	    \begin{split}
	    \sigma_N=&\frac{(\lambda_{S\chi_s}^{\prime})^2 (c^S_G)^2 m_N^2\mu_N^2 v^2}{8\pi\Lambda^2 m_S^4 m_u^2}\\
		&\cdot\left(\sum_{q=u,d,s}(c_G^q f^q_N)+\frac{2}{9}c_G^g f^g_N\right)^2,
	    \end{split}
	\end{equation}
	and the resulting bounds are shown in Fig.~\ref{fig:cGbounds}.
	For the benchmark scenarios $(m_S=400\, \mathrm{GeV}, m_{\chi_s}=500\, \mathrm{GeV})$ and 
	$(m_S=500\, \mathrm{GeV}, m_{\chi_s}=1\, \mathrm{TeV})$,  the limits read
	\begin{equation}
	    \label{eq:limq}
	    \begin{split}
		&\frac{c_G^S}{\Lambda}|_{m_{\chi_s}=500\, \mathrm{GeV}}
		    \lesssim  310 f_{\mathrm{rel}}^{-1/4}\, \mathrm{TeV}^{-1},\\
		&\frac{c_G}{\Lambda}|_{m_{\chi_s}=1\, \mathrm{TeV}}
		    \lesssim  780 f_{\mathrm{rel}}^{-1/4}\, \mathrm{TeV}^{-1}.
	    \end{split}
	\end{equation}

	In consequence, the monojet cross section
	is bounded by
	\begin{equation}
	    \begin{split}
	    &\sigma_j|_{m_{\chi_s}=500\,\mathrm{GeV}} \lesssim 670\, \mathrm{fb},\\ 
	    &\sigma_j|_{m_{\chi_s}=1\,\mathrm{TeV}} \lesssim 140\, \mathrm{fb}. 
	    \end{split}
	\end{equation}
	Again the cross sections are comparable to the fermionic-DM case, and the conclusion that the gluon-induced production provides an 
	interesting scenario also for collider searches.

    \subsection{Higgs--mediator portal}

\begin{figure}[!t]
	\begin{center}
	    \includegraphics[height=2.6in]{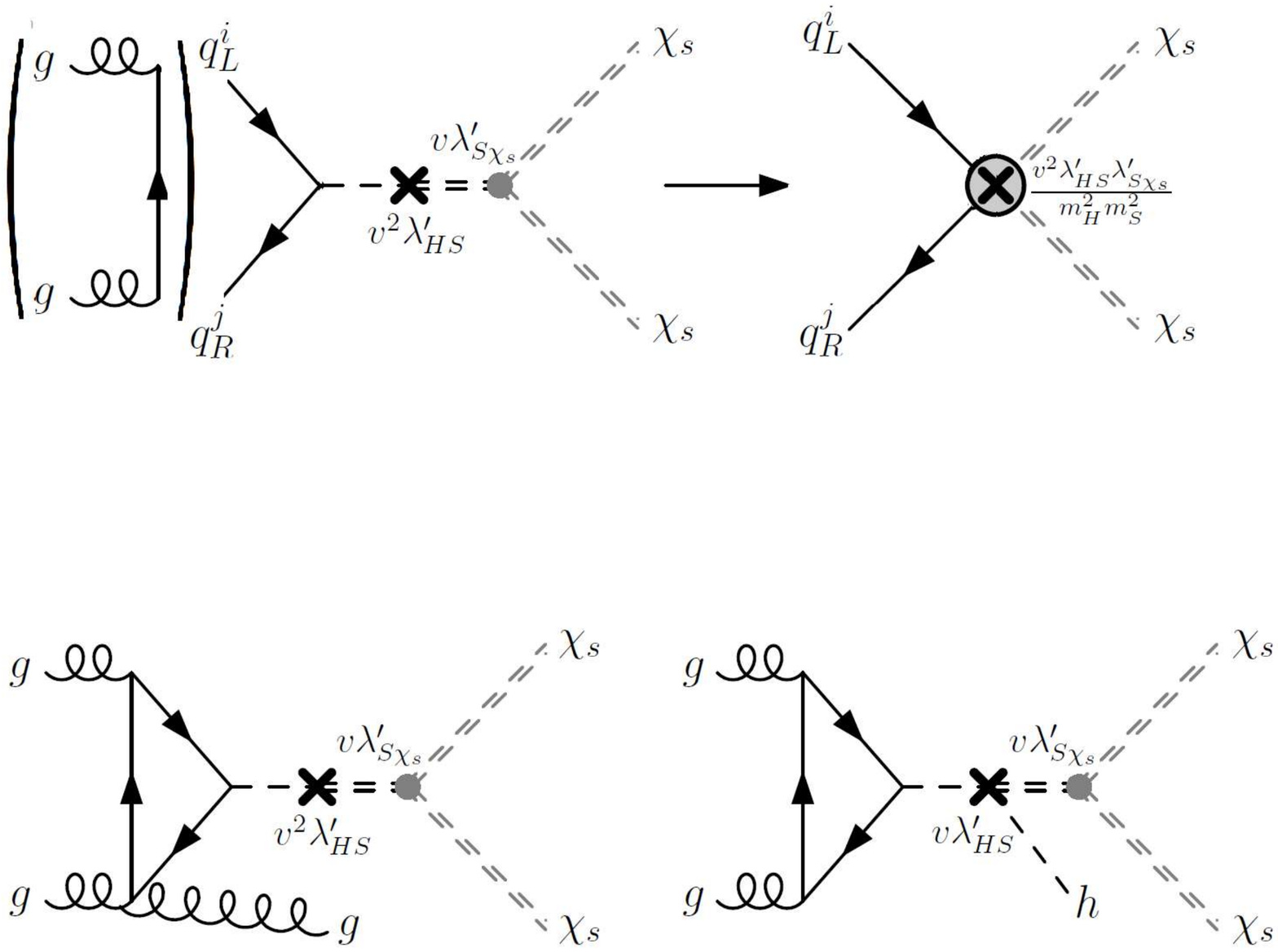} 
	    \caption{\label{fig:diagsS4} Relevant diagrams contributing to nuclear interaction with scalar DM (first row)
		 and corresponding DM observables at hadron colliders (monojet and Higgs$+\MET$, second row), turning
		 on the interactions $\sim \lambda_{H S}^\prime$ and $\sim \lambda_{S \chi_s}^\prime$.}
	\end{center}
    \end{figure}

    Another interesting option is to couple the DM to the SM via the portal 
    term involving the mediator and the Higgs field, $\sim \lambda_{HS}^\prime$,
    turning on this coupling in addition to $y_S$ ($\lambda_{S \chi_s}^\prime$)
    for fermionic (scalar) DM.
    This allows for a SM-like production of the scalar mediator via its mixing with the Higgs field, 
    while its coupling to the DM remains as in the cases discussed above.
    The relevant diagrams are given in Figs~\ref{fig:diags4} and~\ref{fig:diagsS4}. 
    Note that in the figures for simplicity, we employ a mass-insertion approximation,
    where the mixing of the Higgs with the new scalar is treated as an interaction,
    marked by a black cross. In the numerical analysis below, we instead 
    diagonalize the $H-{\cal S}$ system.

    Moreover, the Higgs$+\MET$ channel receives the 
    contribution from the second Higgs field present in the portal,
    as depicted by the lower-right diagrams in Figs~\ref{fig:diags4} and~\ref{fig:diagsS4},
    respectively.

    We diagonalise the Higgs--${\cal S}$ system via a rotation
    \begin{equation}
	\label{eq:mixing}
	\begin{split}
	    h^0=& h \cos\alpha +{\cal S} \sin\alpha,\\
	    H^0=&- h \sin\alpha +{\cal S} \cos\alpha,
	\end{split}
    \end{equation}
    with the angle, $\alpha$, given by
    \begin{equation}
	\label{eq:mixingAngle}
	\tan 2\alpha=\frac{2\lambda_{HS}^{\prime}v^2}{2\lambda_H v^2-\mu_S^2}.
    \end{equation}
    Further, we trade $\lambda_H$ (the coefficient of the quartic Higgs operator) and $\mu_S^2$ for the masses of the eigenstates, $m_h$ and $m_H$, the lighter of which we 
    identify with the 125-GeV Higgs boson.

	\subsubsection{Fermionic DM}

	For fermionic DM, we finally obtain
	\begin{equation}
	    \label{eq:}
	    \begin{split}
	    \sigma_N=&\frac{y_S^2 f_N^2 m_N^2\mu_N^2}{\pi v^2}\sin^2\alpha\cos^2\alpha\left(\frac{1}{m_h^2}-\frac{1}{m_H^2}\right)^2\\
	    =&\frac{y_S^2 f_N^2 m_N^2\mu_N^2(\lambda_{HS}^{\prime})^2v^2}{\pi m_h^4 m_H^4}.
	    \end{split}
	\end{equation}
	For the relic density calculation, we now add the $\bar{t}t$ channel mediated by the Higgs doublet. We focus on $m_{\chi}>m_t$ and
	use the estimate~\cite{Arcadi:2017kky}
	\begin{equation}
	    \begin{split}
		\label{eq:frelHportfDM}
		\langle\sigma \vel\rangle(\bar{\chi}\chi\rightarrow \bar{t}t)
		\approx& 4.0\times 10^{-26}\, \mathrm{cm}^3 \mathrm{s}^{-1}\\
		&\cdot y_S^2\sin^2\alpha\cos^2\alpha\left(\frac{1\, \mathrm{TeV} }{m_{\chi}}\right)^2.  
	    \end{split}
	\end{equation}
	For the scalar channels, we use Eq.~\eqref{eq:frelfDM} after scaling the couplings with the appropriate mixing coefficients given in
	Eq.~\eqref{eq:mixing}.
	
\begin{figure}
	    \begin{center}
		\includegraphics[width=0.45\textwidth]{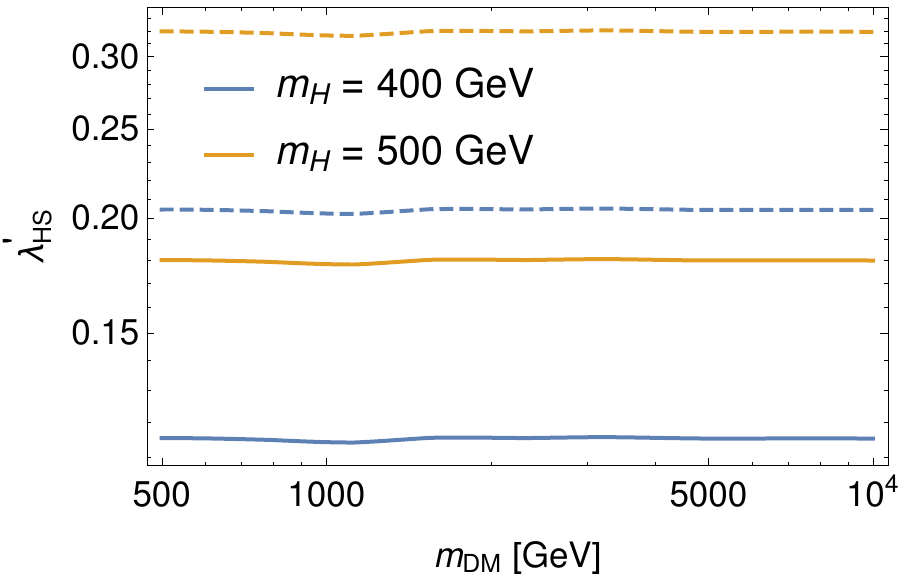}
	    \end{center}
	    \caption{Limits from the direct-detection experiments for the coupling $\lambda_{HS}^{\prime}$ as a function of the DM mass for the 
	    Higgs--mediator portal scenario with fermionic DM.  
	    Blue (yellow) lines correspond to a mass of the heavy scalar eigenstate of $400$~GeV (500~GeV). Solid (dashed) lines correspond to 
	    $f_{\mathrm{rel}}=1$ ($f_{\mathrm{rel}}=0.1$).}
	    \label{fig:mixfbounds}
	\end{figure}

	Comparing now with XENON1T results, leads to the  bounds shown in Fig.~\ref{fig:mixfbounds},
	and the monojet 
	cross section
	is bounded due to the direct detection limit as
	\begin{equation}
	    \begin{split}
	    &\sigma_j|_{m_{\chi}=500\,\mathrm{GeV}} \lesssim 1.1\cdot 10^{-3}\, \mathrm{fb},\\ 
	    &\sigma_j|_{m_{\chi}=1\,\mathrm{TeV}} \lesssim 3.3\cdot 10^{-4}\, \mathrm{fb}. 
	    \end{split}
	\end{equation} 

	\subsubsection{Scalar DM}

\begin{figure}
	    \begin{center}
		\includegraphics[width=0.45\textwidth]{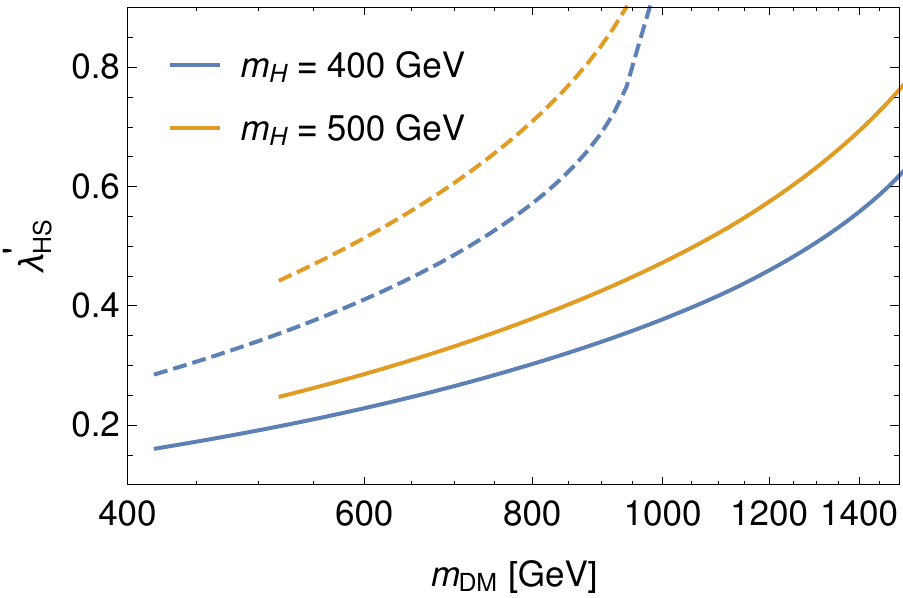}
	    \end{center}
	    \caption{Limits from the direct-detection experiments for the coupling $\lambda_{HS}^{\prime}$ as a function of the DM mass for 
	    the Higgs--mediator portal scenario with scalar DM.  
	    Blue (yellow) lines correspond to a mass of the heavy scalar eigenstate of $400$~GeV (500~GeV). Solid (dashed) lines correspond to 
	    $f_{\mathrm{rel}}=1$ ($f_{\mathrm{rel}}=0.1$).}
	    \label{fig:mixSbounds}
	\end{figure}

	For scalar DM, we arrive at
	\begin{equation}
	    \label{eq:}
	    \begin{split}
	    \sigma_N=&\frac{(\lambda_{S\chi}^{\prime})^2 f_N^2 m_N^2\mu_N^2}{8\pi m_{\chi_s}^2}\sin^2\alpha\cos^2\alpha
		\left(\frac{1}{m_h^2}-\frac{1}{m_H^2}\right)^2\\
	    =&\frac{(\lambda_{S\chi_s}^{\prime})^2 f_N^2 m_N^2\mu_N^2(\lambda_{HS}^{\prime})^2v^4}{8\pi m_{\chi_s}^2 m_h^4 m_H^4}.
	    \end{split}
	\end{equation}
	The annihilation cross section to top quarks for ${m_{\chi_s}>m_t}$ now becomes~\cite{Arcadi:2017kky}
	\begin{equation}
	    \begin{split}
		\label{eq:frelHportfDM}
		\langle\sigma \vel\rangle(\chi_s\chi_s\rightarrow \bar{t}t)
		\approx& 2.6\times 10^{-27}\, \mathrm{cm}^3 \mathrm{s}^{-1}\\
		&\cdot \lambda_{S\chi_s}^{\prime\,2}\sin^2\alpha\cos^2\alpha\left(\frac{1\, \mathrm{TeV} }{m_{\chi_s}}\right)^4,  
	    \end{split}
	\end{equation}
	and for the scalar channels we use Eq,~\eqref{eq:frelSDM} scaling the couplings with the mixing coeffients, see Eq.~\eqref{eq:mixing}.
	The direct detection bounds are shown in Fig.~\ref{fig:mixSbounds}, and the corresponding LHC cross sections
	are now predicted to be below
	\begin{equation}
	    \begin{split}
	    &\sigma_j|_{m_{\chi_s}=500\,\mathrm{GeV}} \lesssim 4.3\cdot 10^{-4}\, \mathrm{fb},\\ 
	    &\sigma_j|_{m_{\chi_s}=1\,\mathrm{TeV}} \lesssim 1.7\cdot 10^{-4}\, \mathrm{fb}\,.
	    \end{split}
	\end{equation}
	Although this process is not observable in this restricted scenario, the portal contribution 
	could become relevant in a combined analysis with more operators turned on.

    \subsection{Higgs--DM portal}

    Finally, the DM could also be coupled {\it directly} to the SM via the 
    Higgs--DM portal,
    promoting the SM-like Higgs boson itself to a mediator to the 
    DM sector, simply by turning on the $D=5$ operator 
    $\sim y_H^{(2)}$ or $D=4$ portal $\sim \lambda_{H \chi_s}$, for
    fermionic or scalar DM, respectively. This corresponds to 
    the traditional SM+singlet DM model studied extensively in the literature; see e.g. 
    Refs~\cite{McDonald:1993ex,Burgess:2000yq,Cline:2013gha,LopezHonorez:2012kv,Alanne:2014bra}.

\begin{figure}[!t]
	\begin{center}
	    \includegraphics[height=2.6in]{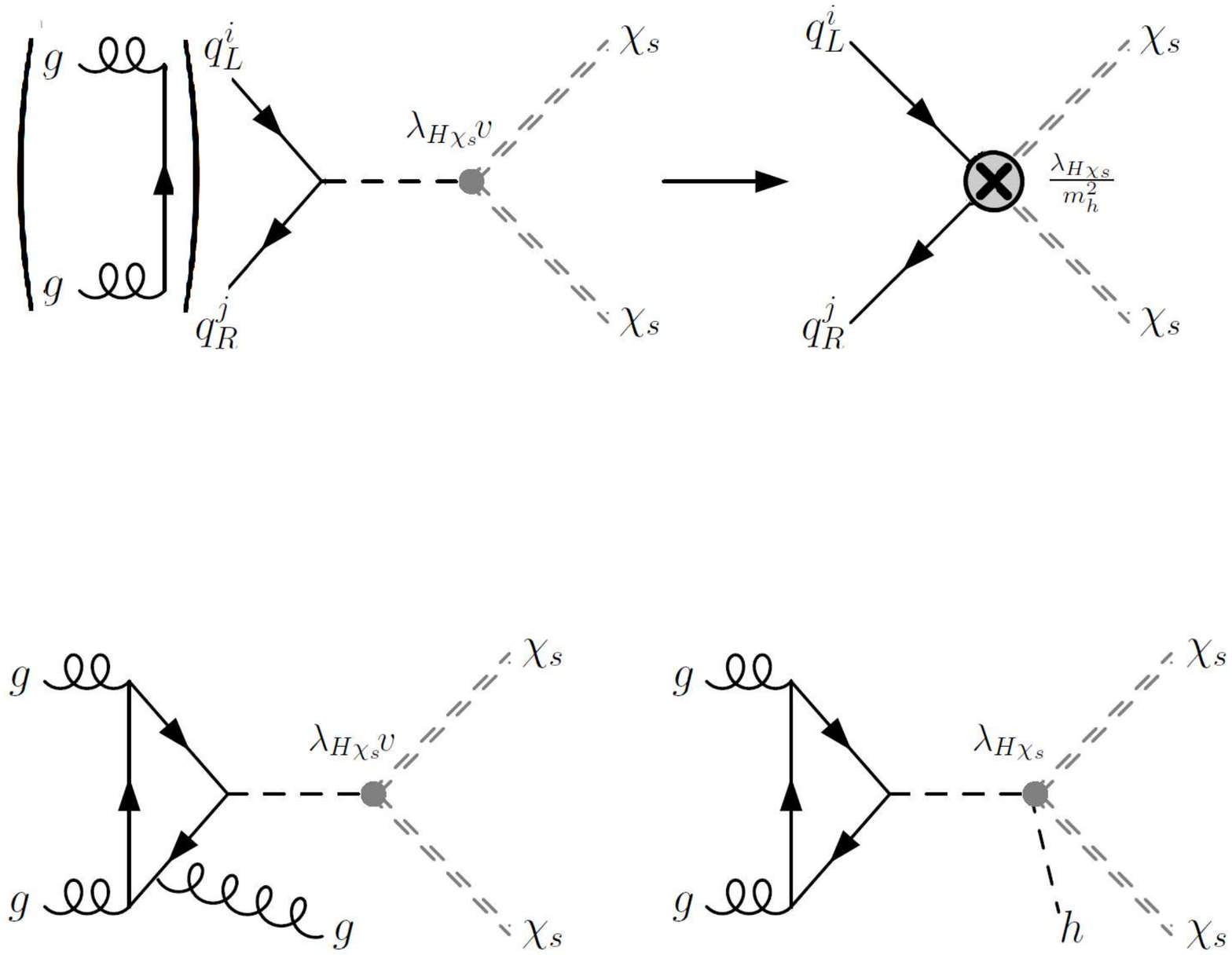} 
	    \caption{\label{fig:diagsS5} Relevant diagrams contributing to nuclear interaction with scalar DM (first row)
		 and corresponding DM observables at hadron colliders (monojet and Higgs$+\MET$, second row), turning
		 on the interaction $\sim \lambda_{H \chi_s}$.}
	\end{center}
    \end{figure}

    While the invisible width of the Higgs boson significantly constrains
    these operators in case the DM is light, $m_{\chi_{(s)}} < m_h/2$
    (see below), for heavier DM they might play an important role in 
    DM production. The corresponding Feynman diagrams for the 
    processes at hand are given in Figs~\ref{fig:diags5} and~\ref{fig:diagsS5}.
    Again, the Higgs$+\MET$ channel is induced at leading order via the second 
    Higgs field required by gauge invariance, see the lower-right diagrams.

	\subsubsection{Fermionic DM}
\begin{figure}
	    \begin{center}
		\includegraphics[width=0.45\textwidth]{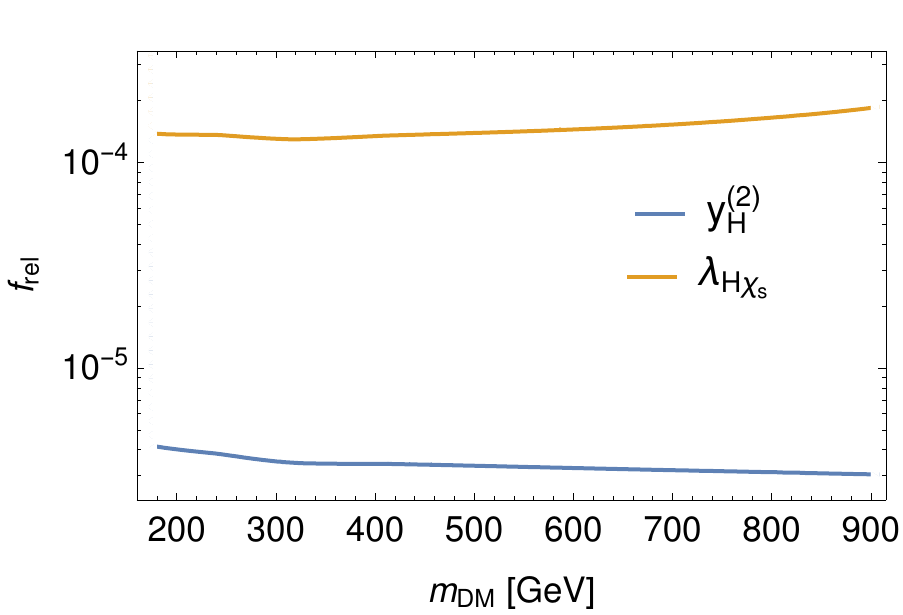}
	    \end{center}
	    \caption{The blue (yellow) curve shows the maximum allowed fraction of the DM abundance, $f_{\mathrm{rel}}$, as a 
	    function of the DM mass fixed by limits from direct detection for the Higgs--DM portal scenario with fermionic (scalar) DM. We only show 
	    the mass range $m_{\mathrm{DM}}\lesssim 900$~GeV where the relic abundance is reproduced for
	    $\lambda_{H\chi_{s}}<4\pi$.} 
	    \label{fig:Hportlim}
	\end{figure}

	For fermionic DM, we obtain the direct detection cross section
	\begin{equation}
	    \label{eq:}
	    \sigma_N=\frac{(y_H^{(2)})^2 f_N^2 m_N^2\mu_N^2}{\pi \Lambda^2m_h^4}.
	\end{equation}
	We trade $y_H^{(2)}/\Lambda$ for $f_{\mathrm{rel}}$ from the thermal $\bar{\chi}\chi\rightarrow \bar t t, WW, ZZ, hh$
	cross section for fixed $m_{\chi_s}$. The dominant contribution corresponds to the $t$-channel annihilation to $hh$, which
	can be obtained using Eq.~\eqref{eq:frelfDM} with substitution 
	$y_S\rightarrow y_H^{(2)}v/\Lambda$.
	We show the direct detection limits in the $(m_{\mathrm{DM}},f_{\mathrm{rel}})$ plane in Fig.~\ref{fig:Hportlim}. 
	The monojet 
	cross section 
	is in turn limited to
	\begin{equation}
	    \begin{split}
	    &\sigma_j|_{m_{\chi}=500\,\mathrm{GeV}} \lesssim 49\, \mathrm{fb},\\ 
	    &\sigma_j|_{m_{\chi}=1\,\mathrm{TeV}} \lesssim 3.3\, \mathrm{fb}.\\ 
	    \end{split}
	\end{equation}
	This scenario could provide another interesting probe for the collider experiments, but is constrained to deliver only
	a tiny fraction of the DM relic abundance.

	\subsubsection{Scalar DM}
	For the thermal cross section, including $\chi_s\chi_s \to \bar t t, WW, ZZ, hh $, we use the result from~\cite{Cline:2012hg}:
	\begin{equation}
	    \label{eq:}
	    \begin{split}
		\langle\sigma \vel\rangle=&\frac{\lambda_{H\chi_s}^2}{2\pi m_{\chi_s}^2(4-r_h)^2}\left[
		6r_t(1-r_t)^{3/2}\right.\\
		&+\sum_{V=W,Z}\delta_V r_V^2(2+(1-2/r_V)^2)\sqrt{1-r_V}\\
		&+\left. 2\left(\frac{2\lambda_{H\chi_s}}{\lambda_H}\frac{(1-r_h/4)r_h}{r_h-2}+1 + \frac{r_h}{2}\right)^2\sqrt{1-r_h}
		\right],
	    \end{split}
	\end{equation}
	where $r_i=m_i^2/m_{\chi_s}^2$, and $\delta_W=1,\delta_Z=1/2$. 
	
	The direct detection cross section now reads
	\begin{equation}
	    \label{eq:}
	    \sigma_N=\frac{\lambda_{H\chi_s}^2\mu_N^2m_N^2f_N^2}{\pi m_{\chi_s}^2m_h^4},
	\end{equation}
	so we plot the direct detection limits in the  $(m_{\mathrm{DM}},f_{\mathrm{rel}})$ plane in Fig.~\ref{fig:Hportlim}.
	The LHC cross section in that final case
	is thus constrained to
	\begin{equation}
	    \begin{split}
	    &\sigma_j|_{m_{\chi_s}=500\,\mathrm{GeV}} \lesssim 0.38\, \mathrm{fb}.\\ 
	    \end{split}
	\end{equation}
	We note that the cross section is somewhat smaller than that of the fermionic-DM case, making the latter scenario more promising for the collider experiments.

    \subsection{Further Processes}

    While dedicated analyses are left for future work, here we already comment on further potentially 
    interesting applications of the \eDMEFT\ framework ranging from the inclusion of loop processes
    in the EFT, over new production mechanisms, up to analyses of invisible Higgs decays, Higgs 
    pair production and collider searches for the mediator.

	\subsubsection{Loop mediated processes in \eDMEFT}

\begin{figure}[!t]
	    \begin{center}
		\includegraphics[height=0.95in]{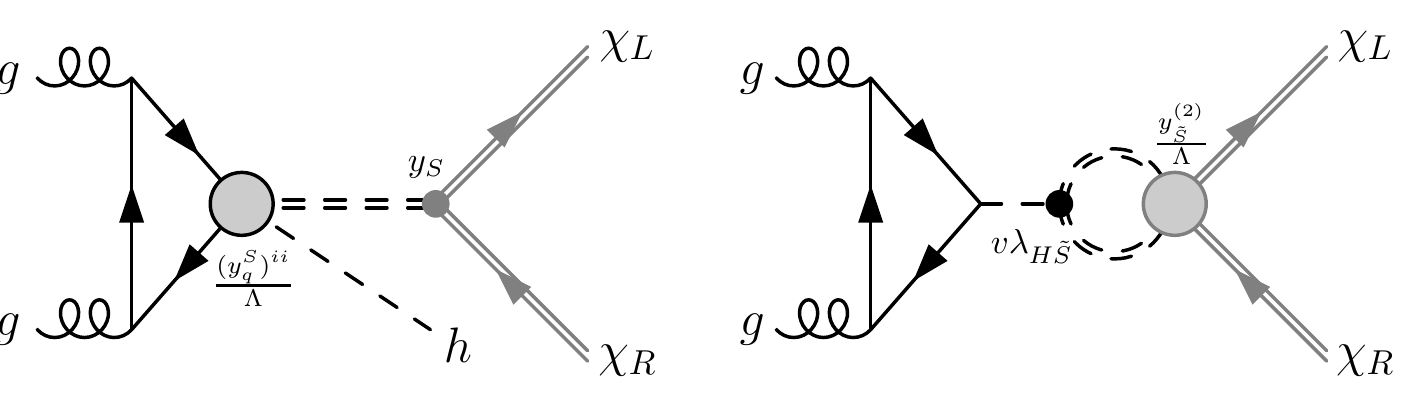} 
		\caption{\label{fig:diagsloop} Potentially important loop diagrams 
		    contributing to DM interactions in the \eDMEFT, turning on the operators $\sim (y_q^S)^{ij}$
		    and $\sim y_S$, or $\sim \lambda_{H \tilde S}$ and $\sim y_{\tilde S}^{(2)}$, respectively. The right diagram
		    involves the quartic portal term $\sim \lambda_{H \tilde S}$, which is basically the only relevant term that allows to
		    produce scalar DM via a (pair-produced) pseudo-scalar mediator at the $D\leq 5$ level (via a similar diagrams with 
		    $y_{\tilde S}^{(2)}/\Lambda \to \lambda_{\tilde S \chi_s}$). 
		    See text for details.}
	    \end{center}
	\end{figure}

	While the only loop diagrams we encountered so far contained the SM-like $GG h$ triangle, loops
	involving ${D=5}$ vertices allow for interesting new means to couple the DM to hadrons.
	First of all, the Yukawa-like operator $\sim (y_q^S)^{ij}$ can now be inserted coming with 
	the top quark ($q=u;\,ij=33$) and coupling the mediator to a gluon pair via a (top-)quark
	loop, see the left diagram in Fig.~\ref{fig:diagsloop}. This operator is expected to be sizable, 
	featuring no (minimal-flavour-violation-like) flavour suppression. The loop suppression is thus lifted by the expected 
	enhancement with $m_t/m_q$, since the operator is basically unconstrained for the top.
	A detailed study of flavour constraints on the coefficients, combining them with complementary bounds 
	from limits on the invisible Higgs width and other searches, such as to derive 
	conclusive limits on DM production in the \eDMEFT, will be presented elsewhere.

	Finally, at the two-loop level, DM production via a portal to a {\it pseudo-scalar} mediator becomes
	possible, see the right diagram in Fig. \ref{fig:diagsloop}. Applied to the case of scalar DM
	(i.e. replacing $y_{\tilde S}^{(2)}/\Lambda  \to \lambda_{\tilde S \chi_s}$), this opens the possibility
	to obtain potentially viable models with a CP-odd mediator.

	\subsubsection{Weak-boson fusion}

\begin{figure}[!t]
    \begin{center}
	\includegraphics[height=3.75in]{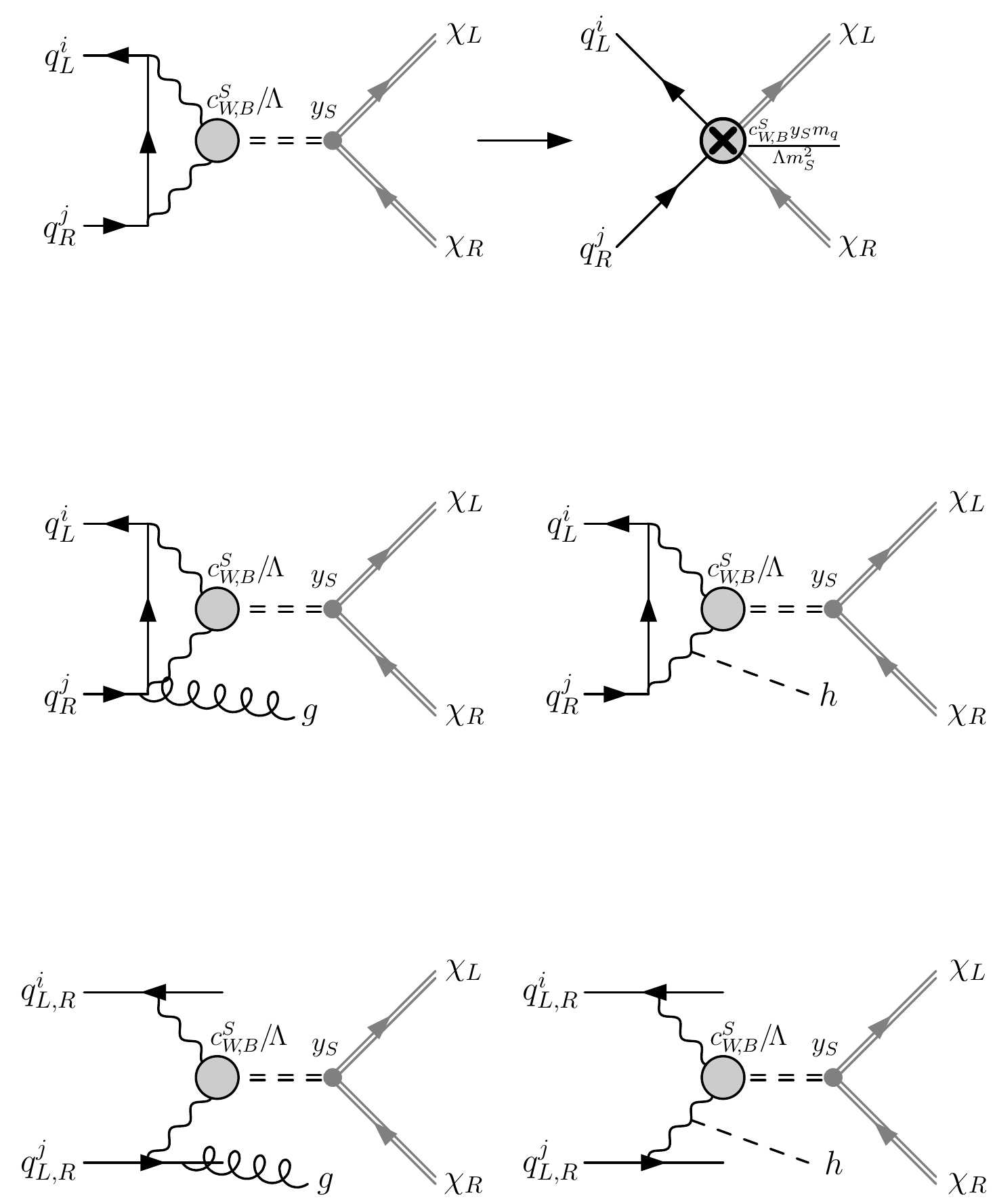} 
	\caption{\label{fig:diags3}  Diagrams contributing to nuclear interaction 
	    with fermionic DM (first row) and DM observables at hadron colliders, turning
	    on the interactions $\sim c_W^S$ and $\sim y_S$. 
	    The diagrams are similar for pseudo-scalar mediators, employing the corresponding 
	    tilded coefficients, as discussed before. See text for details. }
    \end{center}
\end{figure}

\begin{figure}[!t]
	    \begin{center}
		\includegraphics[height=3.75in]{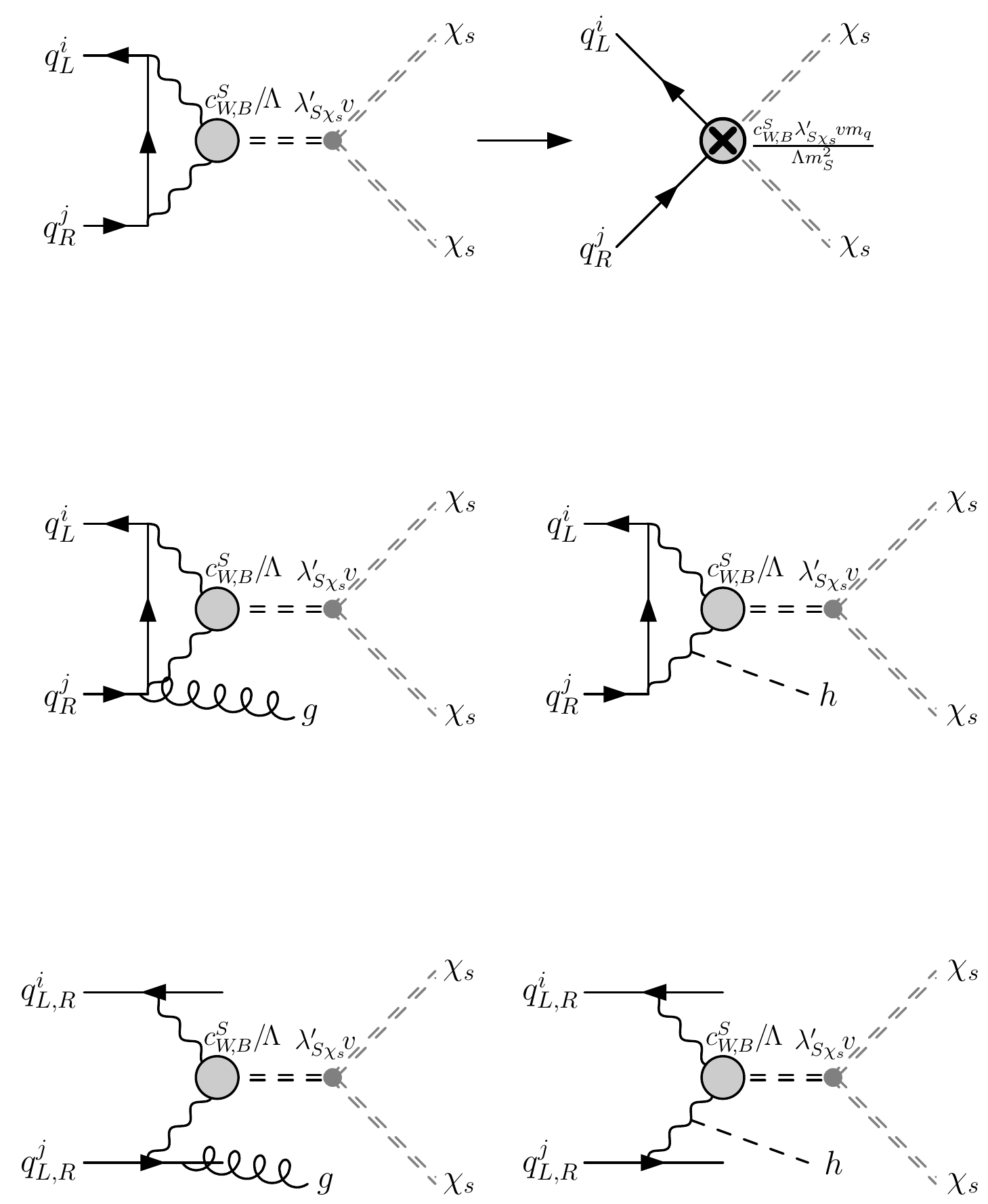} 
		\caption{\label{fig:diagsS3} Diagrams contributing to nuclear interaction 
		    with scalar DM (first row) and DM observables at hadron colliders, turning
		    on the interactions $\sim c_W^S$ and $\sim \lambda_{S \chi_s}^\prime$. 
		    See text for details.
}
	    \end{center}
	\end{figure}

	In the same context, a production of the mediator in weak-boson fusion, as depicted by the
	last two diagrams in Figs \ref{fig:diags3} and \ref{fig:diagsS3}, is interesting regarding 
	`monojet' and Higgs$+\MET$ signals. In fact, if the corresponding 
	$D=5$ operators feature a sizable coupling $c_{W,B}^S$ (or $S\to \tilde S$, for a CP-odd mediator), DM signals at the LHC
	can be significant, while limits from direct-detection experiments are met, since the corresponding processes 
	with external quark bi-linears are suppressed by a quark-mass insertion (and a loop factor), see the first (and second) 
	row of Figs \ref{fig:diags3} and \ref{fig:diagsS3}.

	\subsubsection{Higgs pair $+\MET$}

\begin{figure}[!t]
	    \begin{center}
	    	\includegraphics[height=2.56in]{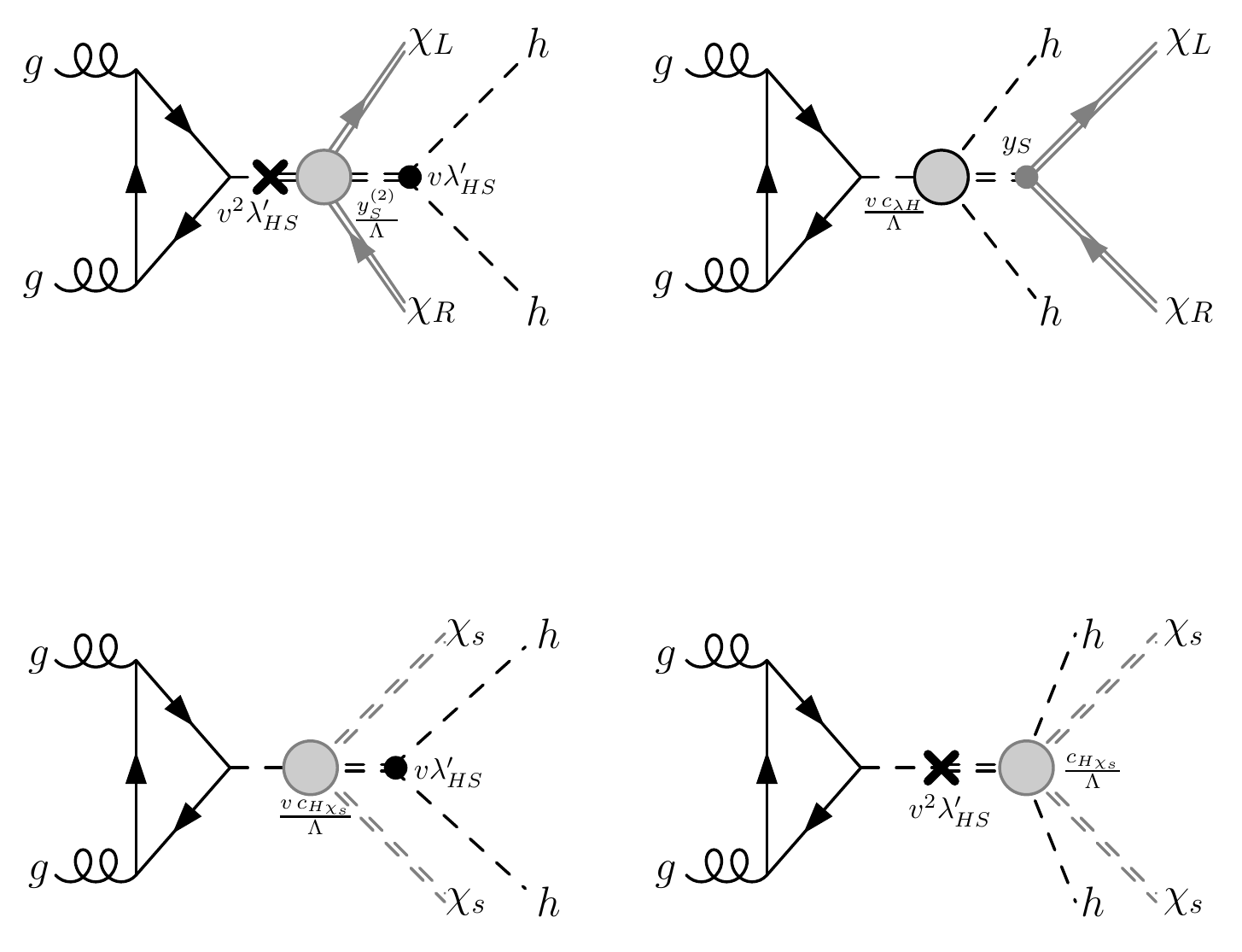} 
		\caption{\label{fig:diagshh} Selected diagrams  
		    contributing to Higgs pair production in association with $\MET$, turning on the interactions 
		    $\sim \lambda_{H S}^\prime, y_{S}^{(2)}$ or $\sim c_{\lambda H}, y_S$
		    in the case of fermionic DM (first row) and
		    $\sim \lambda_{H S}^\prime, c_{H \chi_s}$ in the case of scalar DM (second row).
		    See text for details.}
	    \end{center}
	\end{figure}

	Exploring the production of Higgs pairs in association with missing energy could be an additional interesting
	probe of dark sectors. Indeed, several operators in the \eDMEFT\ allow the production of Higgs pairs
	along with DM and can be tested in this process.
	Fig.~\ref{fig:diagshh} shows sample diagrams that become potentially important in case the 
	bi-quadratic portal $\sim \lambda_{HS}$ is weak. Beyond the usual case of the mediator
	decaying to the DM, they also contain potentially resonant decays of ${\cal M}$ to a Higgs pair,
	after emission of a DM pair, which could allow to see a peak in the $hh$ invariant mass
	spectrum and boosted Higgs bosons \cite{Kang:2015nga}
	\footnote{Note that, for the given couplings turned on,
	there are additional (potentially) similar important contributions, attaching the Higgs or 
	mediator lines differently in the diagrams above (including Higgs emissions from SM lines).}. 
	Although the cross sections are not expected to be large, non-negligible NP couplings could 
	still feature interesting effects in $hh+\MET$ production. A dedicated analysis is needed to 
	examine the actual prospects of this process in the light of the expected limited number of events. 

	\subsubsection{Invisible Higgs decays}

\begin{figure}[!t]
	    \begin{center}
		\includegraphics[height=2.75in]{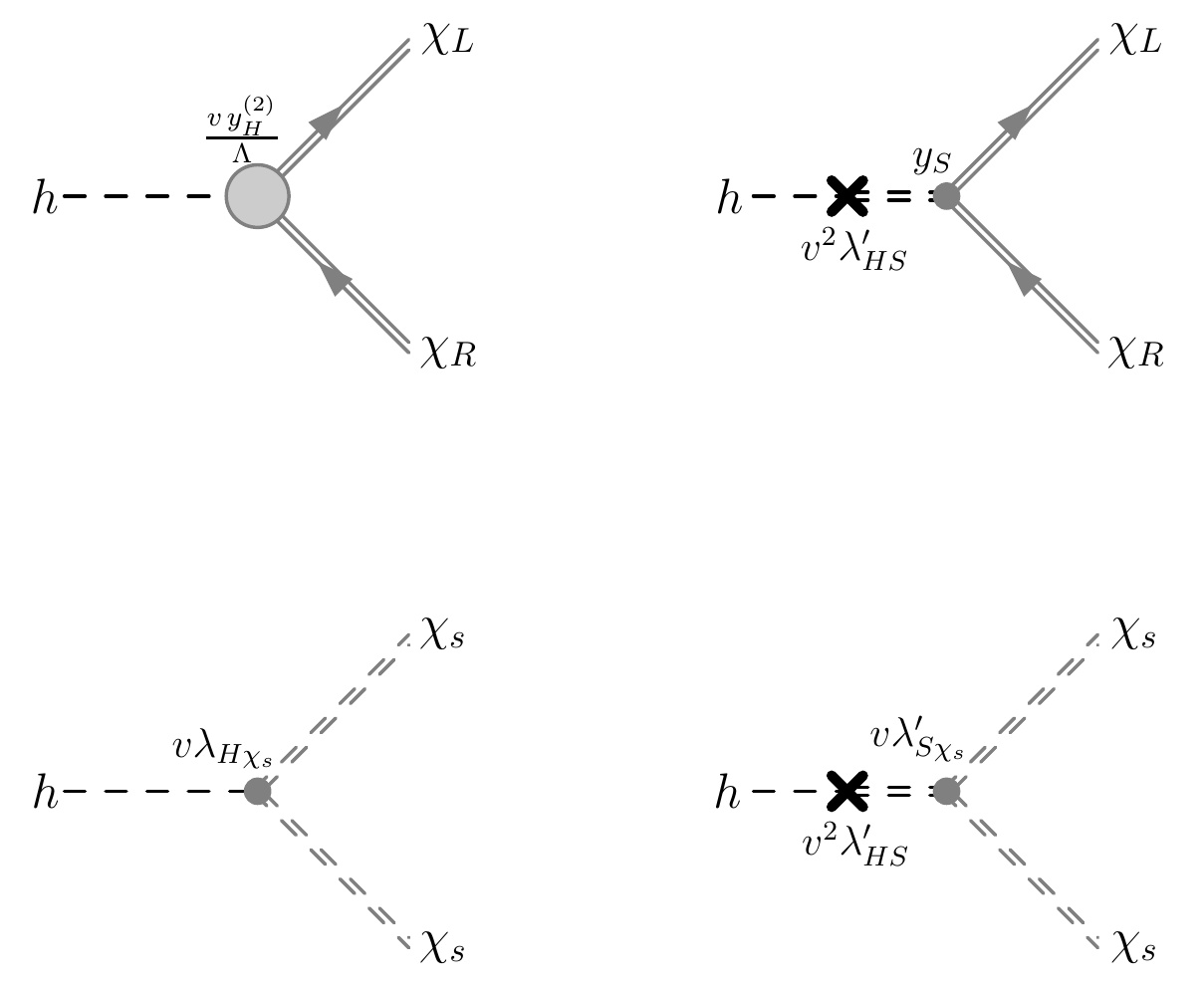} 
		\caption{\label{fig:diagsinv} Contributions to the invisible width of the Higgs boson from turning on
		    the interactions
		    $\sim\!y_{H}^{(2)}$, $\sim\!\lambda_{H S}^\prime, y_S$, 
		    $\sim\!\lambda_{H \chi_s}$, $\sim\!\lambda_{H S}^\prime, \lambda_{S \chi_s}^\prime$.}
	    \end{center}
	\end{figure}

	Note that the \eDMEFT\ can also significantly effect single Higgs physics.
	For example, for light DM ($m_{\cal D}\leq m_h$) the \eDMEFT\ operators involving the Higgs field 
	(including $D=4$ Higgs portals) are constrained more and more severely from limits on 
	invisible Higgs decays. Corresponding diagrams, involving the portal coupling
	$\lambda_{H S}^\prime$ together with $y_S$ or $\lambda_{S \chi_s}^\prime$ for the 
	decay to DM are given in the right-hand side of Fig.~\ref{fig:diagsinv}, while those
	with direct $h\!-\!{\cal D}\!-\!{\cal D}$ interactions $y_{H}^{(2)}$ or $\lambda_{H \chi_s}$ are depicted
	in the left-hand side of the same figure. A production of the mediator via 
	$D=5$ \eDMEFT\ operators avoiding these constraints becomes particularly interesting.

	Moreover, the additional scalar particle 
	${\cal S}$ can have an interesting impact on 
	the nature of the electroweak phase transition, 
	which in the SM is not first order, 
	such as to allow for electroweak baryogenesis \cite{Kajantie:1996mn, Rummukainen:1998as}.
	With the help of a light ${\cal S}$, electroweak baryogenesis can 
	become viable (see, e.g. \cite{Anderson:1991zb,Espinosa:1993bs,Choi:1993cv,McDonald:1993ey}).
	In turn, the DM phenomenology will be affected.
	Including such a scenario in our framework is also left for the future.

	\subsubsection{${\cal S}$ Resonance search}

\begin{figure}[!t]
	    \begin{center}
		\includegraphics[height=1.in]{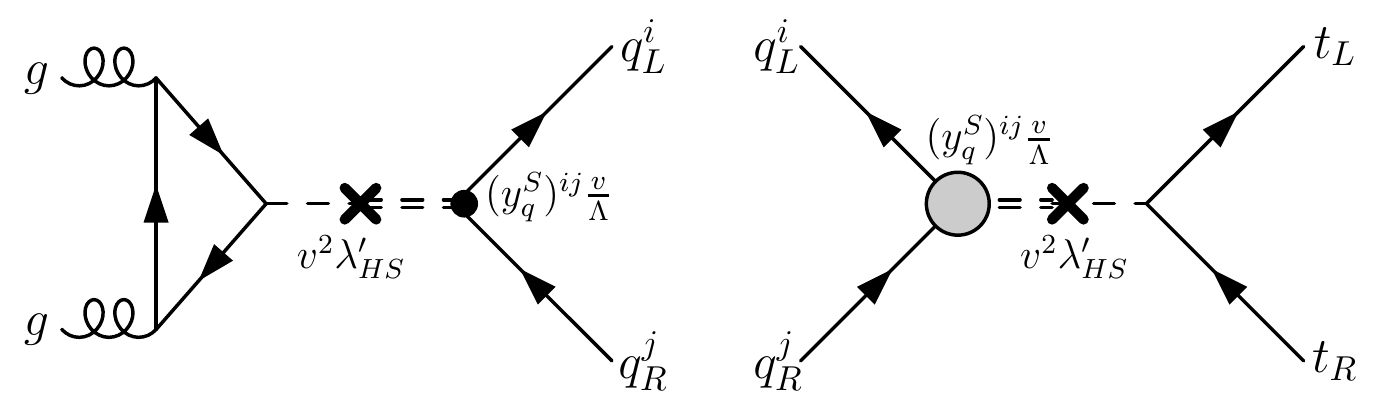} 
		\caption{\label{fig:diagsres} 
		    Resonant contributions of the mediator to di-jet and $t \bar t$ final states,
		    turning on the interactions $\sim \lambda_{H S}^\prime$ and $\sim (y_q^S)^{ij}$.
		    Replacing the latter by $c_{W,B,G}^S$ or $(y_\ell^S)^{ij}$ leads to further interesting
		    contributions to di-boson, di-jet, and di-lepton final states.}
	    \end{center}
	\end{figure}

	Finally, one can also search directly for the mediator by looking for a resonant enhancement in the
	di-jet, $t \bar t$, di-lepton, or di-boson spectrum. These processes can also be described
	consistently in the \eDMEFT, since the inclusion of the mediator as a dynamical degree of freedom
	is a defining feature of the setup, and their study can deliver important insight on the nature of DM.
	Sample diagrams for the first two processes, involving the couplings $\lambda_{H S}^\prime$
	and $(y_q^S)^{ij}$ are shown in Fig. \ref{fig:diagsres}, and the other resonant 
	processes are generated via similar diagrams (including the production of the
	resonance in gluon fusion and the decay via Higgs mixing). Again, the black cross denotes a mass insertion, 
	while in the (diagonal) mass basis just the heavy mediator is exchanged in the s-channel, with its 
	coupling to the $t \bar t$ state governed by the Higgs admixture. 
	Resonance searches are in fact a powerful complementary tool to understand dark sectors
	and to probe the coupling structure of the mediator even in the case where its production
	is dominated by a single operator. Being able to combine this information with that of the other DM observables 
	in a single consistent, yet general, framework is a particular strength of the \eDMEFT.

\section{Conclusions}

We have presented a new framework to describe DM phenomenology,
which can be used to consistently confront limits from direct-detection 
experiments and the relic abundance with possible collider signatures
at high energies, while maintaining a high degree of model-independence.
Both the DM and mediator fields remain 
as propagating degrees of freedom, whereas additional new physics
is described in the form of higher-dimensional operators.
As an application of the framework, focusing on the level of 
$D=5$ operators, we derived possible cross sections in monojet 
and Higgs$+E_{\rm miss}$ signatures at the LHC. Considering the limits
from direct detection and reproducing (a certain fraction of) the
DM relic density, for scalar mediator, we found that the most promising case for the collider searches is
gluon induced production, which could lead to sizable LHC cross sections both for fermionic and scalar DM. 
Cases with pseudoscalar mediators are still rather unconstrained by direct-detection experiments, 
and \eDMEFT\ will reveal its full strength with the future generation of direct-detection experiments.
These results are valid
in a more general context than those derived in conventional
(simplified) models.
A plethora of possible further applications of the framework is left for
future work.

\section*{Acknowledgments} 

We are grateful to Giorgio Arcadi, Martin
Bauer, Farinaldo Queiroz, Valentin Tenorth, and
Stefan Vogel for useful discussions and comments.

\clearpage

\bibliography{DMEFT}

\begin{thebibliography}{41}%
\makeatletter
\providecommand \@ifxundefined [1]{%
 \@ifx{#1\undefined}
}%
\providecommand \@ifnum [1]{%
 \ifnum #1\expandafter \@firstoftwo
 \else \expandafter \@secondoftwo
 \fi
}%
\providecommand \@ifx [1]{%
 \ifx #1\expandafter \@firstoftwo
 \else \expandafter \@secondoftwo
 \fi
}%
\providecommand \natexlab [1]{#1}%
\providecommand \enquote  [1]{``#1''}%
\providecommand \bibnamefont  [1]{#1}%
\providecommand \bibfnamefont [1]{#1}%
\providecommand \citenamefont [1]{#1}%
\providecommand \href@noop [0]{\@secondoftwo}%
\providecommand \href [0]{\begingroup \@sanitize@url \@href}%
\providecommand \@href[1]{\@@startlink{#1}\@@href}%
\providecommand \@@href[1]{\endgroup#1\@@endlink}%
\providecommand \@sanitize@url [0]{\catcode `\\12\catcode `\$12\catcode
  `\&12\catcode `\#12\catcode `\^12\catcode `\_12\catcode `\%12\relax}%
\providecommand \@@startlink[1]{}%
\providecommand \@@endlink[0]{}%
\providecommand \url  [0]{\begingroup\@sanitize@url \@url }%
\providecommand \@url [1]{\endgroup\@href {#1}{\urlprefix }}%
\providecommand \urlprefix  [0]{URL }%
\providecommand \Eprint [0]{\href }%
\providecommand \doibase [0]{http://dx.doi.org/}%
\providecommand \selectlanguage [0]{\@gobble}%
\providecommand \bibinfo  [0]{\@secondoftwo}%
\providecommand \bibfield  [0]{\@secondoftwo}%
\providecommand \translation [1]{[#1]}%
\providecommand \BibitemOpen [0]{}%
\providecommand \bibitemStop [0]{}%
\providecommand \bibitemNoStop [0]{.\EOS\space}%
\providecommand \EOS [0]{\spacefactor3000\relax}%
\providecommand \BibitemShut  [1]{\csname bibitem#1\endcsname}%
\let\auto@bib@innerbib\@empty
\bibitem [{\citenamefont {Bruggisser}\ \emph {et~al.}(2016)\citenamefont
  {Bruggisser}, \citenamefont {Riva},\ and\ \citenamefont
  {Urbano}}]{Bruggisser:2016nzw}%
  \BibitemOpen
  \bibfield  {author} {\bibinfo {author} {\bibfnamefont {S.}~\bibnamefont
  {Bruggisser}}, \bibinfo {author} {\bibfnamefont {F.}~\bibnamefont {Riva}}, \
  and\ \bibinfo {author} {\bibfnamefont {A.}~\bibnamefont {Urbano}},\ }\href
  {\doibase 10.1007/JHEP11(2016)069} {\bibfield  {journal} {\bibinfo  {journal}
  {JHEP}\ }\textbf {\bibinfo {volume} {11}},\ \bibinfo {pages} {069} (\bibinfo
  {year} {2016})},\ \Eprint {http://arxiv.org/abs/1607.02475} {arXiv:1607.02475
  [hep-ph]} \BibitemShut {NoStop}%
\bibitem [{\citenamefont {Busoni}\ \emph {et~al.}(2014)\citenamefont {Busoni},
  \citenamefont {De~Simone}, \citenamefont {Morgante},\ and\ \citenamefont
  {Riotto}}]{Busoni:2013lha}%
  \BibitemOpen
  \bibfield  {author} {\bibinfo {author} {\bibfnamefont {G.}~\bibnamefont
  {Busoni}}, \bibinfo {author} {\bibfnamefont {A.}~\bibnamefont {De~Simone}},
  \bibinfo {author} {\bibfnamefont {E.}~\bibnamefont {Morgante}}, \ and\
  \bibinfo {author} {\bibfnamefont {A.}~\bibnamefont {Riotto}},\ }\href
  {\doibase 10.1016/j.physletb.2013.11.069} {\bibfield  {journal} {\bibinfo
  {journal} {Phys. Lett.}\ }\textbf {\bibinfo {volume} {B728}},\ \bibinfo
  {pages} {412} (\bibinfo {year} {2014})},\ \Eprint
  {http://arxiv.org/abs/1307.2253} {arXiv:1307.2253 [hep-ph]} \BibitemShut
  {NoStop}%
\bibitem [{\citenamefont {Bauer}\ \emph
  {et~al.}(2017{\natexlab{a}})\citenamefont {Bauer}, \citenamefont {Butter},
  \citenamefont {Desai}, \citenamefont {Gonzalez-Fraile},\ and\ \citenamefont
  {Plehn}}]{Bauer:2016pug}%
  \BibitemOpen
  \bibfield  {author} {\bibinfo {author} {\bibfnamefont {M.}~\bibnamefont
  {Bauer}}, \bibinfo {author} {\bibfnamefont {A.}~\bibnamefont {Butter}},
  \bibinfo {author} {\bibfnamefont {N.}~\bibnamefont {Desai}}, \bibinfo
  {author} {\bibfnamefont {J.}~\bibnamefont {Gonzalez-Fraile}}, \ and\ \bibinfo
  {author} {\bibfnamefont {T.}~\bibnamefont {Plehn}},\ }\href {\doibase
  10.1103/PhysRevD.95.075036} {\bibfield  {journal} {\bibinfo  {journal} {Phys.
  Rev.}\ }\textbf {\bibinfo {volume} {D95}},\ \bibinfo {pages} {075036}
  (\bibinfo {year} {2017}{\natexlab{a}})},\ \Eprint
  {http://arxiv.org/abs/1611.09908} {arXiv:1611.09908 [hep-ph]} \BibitemShut
  {NoStop}%
\bibitem [{\citenamefont {Carmona}\ \emph {et~al.}(2017)\citenamefont
  {Carmona}, \citenamefont {Goertz},\ and\ \citenamefont
  {Papaefstathiou}}]{Carmona:2016qgo}%
  \BibitemOpen
  \bibfield  {author} {\bibinfo {author} {\bibfnamefont {A.}~\bibnamefont
  {Carmona}}, \bibinfo {author} {\bibfnamefont {F.}~\bibnamefont {Goertz}}, \
  and\ \bibinfo {author} {\bibfnamefont {A.}~\bibnamefont {Papaefstathiou}},\
  }\href {\doibase 10.1103/PhysRevD.95.095022} {\bibfield  {journal} {\bibinfo
  {journal} {Phys. Rev.}\ }\textbf {\bibinfo {volume} {D95}},\ \bibinfo {pages}
  {095022} (\bibinfo {year} {2017})},\ \Eprint
  {http://arxiv.org/abs/1606.02716} {arXiv:1606.02716 [hep-ph]} \BibitemShut
  {NoStop}%
\bibitem [{\citenamefont {Franceschini}\ \emph {et~al.}(2016)\citenamefont
  {Franceschini}, \citenamefont {Giudice}, \citenamefont {Kamenik},
  \citenamefont {McCullough}, \citenamefont {Riva}, \citenamefont {Strumia},\
  and\ \citenamefont {Torre}}]{Franceschini:2016gxv}%
  \BibitemOpen
  \bibfield  {author} {\bibinfo {author} {\bibfnamefont {R.}~\bibnamefont
  {Franceschini}}, \bibinfo {author} {\bibfnamefont {G.~F.}\ \bibnamefont
  {Giudice}}, \bibinfo {author} {\bibfnamefont {J.~F.}\ \bibnamefont
  {Kamenik}}, \bibinfo {author} {\bibfnamefont {M.}~\bibnamefont {McCullough}},
  \bibinfo {author} {\bibfnamefont {F.}~\bibnamefont {Riva}}, \bibinfo {author}
  {\bibfnamefont {A.}~\bibnamefont {Strumia}}, \ and\ \bibinfo {author}
  {\bibfnamefont {R.}~\bibnamefont {Torre}},\ }\href {\doibase
  10.1007/JHEP07(2016)150} {\bibfield  {journal} {\bibinfo  {journal} {JHEP}\
  }\textbf {\bibinfo {volume} {07}},\ \bibinfo {pages} {150} (\bibinfo {year}
  {2016})},\ \Eprint {http://arxiv.org/abs/1604.06446} {arXiv:1604.06446
  [hep-ph]} \BibitemShut {NoStop}%
\bibitem [{\citenamefont {Gripaios}\ and\ \citenamefont
  {Sutherland}(2016)}]{Gripaios:2016xuo}%
  \BibitemOpen
  \bibfield  {author} {\bibinfo {author} {\bibfnamefont {B.}~\bibnamefont
  {Gripaios}}\ and\ \bibinfo {author} {\bibfnamefont {D.}~\bibnamefont
  {Sutherland}},\ }\href {\doibase 10.1007/JHEP08(2016)103} {\bibfield
  {journal} {\bibinfo  {journal} {JHEP}\ }\textbf {\bibinfo {volume} {08}},\
  \bibinfo {pages} {103} (\bibinfo {year} {2016})},\ \Eprint
  {http://arxiv.org/abs/1604.07365} {arXiv:1604.07365 [hep-ph]} \BibitemShut
  {NoStop}%
\bibitem [{\citenamefont {Luty}(1998)}]{Luty:1997fk}%
  \BibitemOpen
  \bibfield  {author} {\bibinfo {author} {\bibfnamefont {M.~A.}\ \bibnamefont
  {Luty}},\ }\href {\doibase 10.1103/PhysRevD.57.1531} {\bibfield  {journal}
  {\bibinfo  {journal} {Phys. Rev.}\ }\textbf {\bibinfo {volume} {D57}},\
  \bibinfo {pages} {1531} (\bibinfo {year} {1998})},\ \Eprint
  {http://arxiv.org/abs/hep-ph/9706235} {arXiv:hep-ph/9706235 [hep-ph]}
  \BibitemShut {NoStop}%
\bibitem [{\citenamefont {Cohen}\ \emph {et~al.}(1997)\citenamefont {Cohen},
  \citenamefont {Kaplan},\ and\ \citenamefont {Nelson}}]{Cohen:1997rt}%
  \BibitemOpen
  \bibfield  {author} {\bibinfo {author} {\bibfnamefont {A.~G.}\ \bibnamefont
  {Cohen}}, \bibinfo {author} {\bibfnamefont {D.~B.}\ \bibnamefont {Kaplan}}, \
  and\ \bibinfo {author} {\bibfnamefont {A.~E.}\ \bibnamefont {Nelson}},\
  }\href {\doibase 10.1016/S0370-2693(97)00995-7} {\bibfield  {journal}
  {\bibinfo  {journal} {Phys. Lett.}\ }\textbf {\bibinfo {volume} {B412}},\
  \bibinfo {pages} {301} (\bibinfo {year} {1997})},\ \Eprint
  {http://arxiv.org/abs/hep-ph/9706275} {arXiv:hep-ph/9706275 [hep-ph]}
  \BibitemShut {NoStop}%
\bibitem [{\citenamefont {Giudice}\ \emph {et~al.}(2007)\citenamefont
  {Giudice}, \citenamefont {Grojean}, \citenamefont {Pomarol},\ and\
  \citenamefont {Rattazzi}}]{Giudice:2007fh}%
  \BibitemOpen
  \bibfield  {author} {\bibinfo {author} {\bibfnamefont {G.~F.}\ \bibnamefont
  {Giudice}}, \bibinfo {author} {\bibfnamefont {C.}~\bibnamefont {Grojean}},
  \bibinfo {author} {\bibfnamefont {A.}~\bibnamefont {Pomarol}}, \ and\
  \bibinfo {author} {\bibfnamefont {R.}~\bibnamefont {Rattazzi}},\ }\href
  {\doibase 10.1088/1126-6708/2007/06/045} {\bibfield  {journal} {\bibinfo
  {journal} {JHEP}\ }\textbf {\bibinfo {volume} {06}},\ \bibinfo {pages} {045}
  (\bibinfo {year} {2007})},\ \Eprint {http://arxiv.org/abs/hep-ph/0703164}
  {arXiv:hep-ph/0703164 [hep-ph]} \BibitemShut {NoStop}%
\bibitem [{\citenamefont {Chala}\ \emph {et~al.}(2017)\citenamefont {Chala},
  \citenamefont {Durieux}, \citenamefont {Grojean}, \citenamefont {de~Lima},\
  and\ \citenamefont {Matsedonskyi}}]{Chala:2017sjk}%
  \BibitemOpen
  \bibfield  {author} {\bibinfo {author} {\bibfnamefont {M.}~\bibnamefont
  {Chala}}, \bibinfo {author} {\bibfnamefont {G.}~\bibnamefont {Durieux}},
  \bibinfo {author} {\bibfnamefont {C.}~\bibnamefont {Grojean}}, \bibinfo
  {author} {\bibfnamefont {L.}~\bibnamefont {de~Lima}}, \ and\ \bibinfo
  {author} {\bibfnamefont {O.}~\bibnamefont {Matsedonskyi}},\ }\href {\doibase
  10.1007/JHEP06(2017)088} {\bibfield  {journal} {\bibinfo  {journal} {JHEP}\
  }\textbf {\bibinfo {volume} {06}},\ \bibinfo {pages} {088} (\bibinfo {year}
  {2017})},\ \Eprint {http://arxiv.org/abs/1703.10624} {arXiv:1703.10624
  [hep-ph]} \BibitemShut {NoStop}%
\bibitem [{\citenamefont {Goertz}(2017)}]{Goertz:2017gor}%
  \BibitemOpen
  \bibfield  {author} {\bibinfo {author} {\bibfnamefont {F.}~\bibnamefont
  {Goertz}},\ }\href@noop {} {\  (\bibinfo {year} {2017})},\ \Eprint
  {http://arxiv.org/abs/1711.03162} {arXiv:1711.03162 [hep-ph]} \BibitemShut
  {NoStop}%
\bibitem [{\citenamefont {Beltran}\ \emph {et~al.}(2010)\citenamefont
  {Beltran}, \citenamefont {Hooper}, \citenamefont {Kolb}, \citenamefont
  {Krusberg},\ and\ \citenamefont {Tait}}]{Beltran:2010ww}%
  \BibitemOpen
  \bibfield  {author} {\bibinfo {author} {\bibfnamefont {M.}~\bibnamefont
  {Beltran}}, \bibinfo {author} {\bibfnamefont {D.}~\bibnamefont {Hooper}},
  \bibinfo {author} {\bibfnamefont {E.~W.}\ \bibnamefont {Kolb}}, \bibinfo
  {author} {\bibfnamefont {Z.~A.~C.}\ \bibnamefont {Krusberg}}, \ and\ \bibinfo
  {author} {\bibfnamefont {T.~M.~P.}\ \bibnamefont {Tait}},\ }\href {\doibase
  10.1007/JHEP09(2010)037} {\bibfield  {journal} {\bibinfo  {journal} {JHEP}\
  }\textbf {\bibinfo {volume} {09}},\ \bibinfo {pages} {037} (\bibinfo {year}
  {2010})},\ \Eprint {http://arxiv.org/abs/1002.4137} {arXiv:1002.4137
  [hep-ph]} \BibitemShut {NoStop}%
\bibitem [{\citenamefont {Bai}\ \emph {et~al.}(2010)\citenamefont {Bai},
  \citenamefont {Fox},\ and\ \citenamefont {Harnik}}]{Bai:2010hh}%
  \BibitemOpen
  \bibfield  {author} {\bibinfo {author} {\bibfnamefont {Y.}~\bibnamefont
  {Bai}}, \bibinfo {author} {\bibfnamefont {P.~J.}\ \bibnamefont {Fox}}, \ and\
  \bibinfo {author} {\bibfnamefont {R.}~\bibnamefont {Harnik}},\ }\href
  {\doibase 10.1007/JHEP12(2010)048} {\bibfield  {journal} {\bibinfo  {journal}
  {JHEP}\ }\textbf {\bibinfo {volume} {12}},\ \bibinfo {pages} {048} (\bibinfo
  {year} {2010})},\ \Eprint {http://arxiv.org/abs/1005.3797} {arXiv:1005.3797
  [hep-ph]} \BibitemShut {NoStop}%
\bibitem [{\citenamefont {Goodman}\ \emph {et~al.}(2010)\citenamefont
  {Goodman}, \citenamefont {Ibe}, \citenamefont {Rajaraman}, \citenamefont
  {Shepherd}, \citenamefont {Tait},\ and\ \citenamefont {Yu}}]{Goodman:2010ku}%
  \BibitemOpen
  \bibfield  {author} {\bibinfo {author} {\bibfnamefont {J.}~\bibnamefont
  {Goodman}}, \bibinfo {author} {\bibfnamefont {M.}~\bibnamefont {Ibe}},
  \bibinfo {author} {\bibfnamefont {A.}~\bibnamefont {Rajaraman}}, \bibinfo
  {author} {\bibfnamefont {W.}~\bibnamefont {Shepherd}}, \bibinfo {author}
  {\bibfnamefont {T.~M.~P.}\ \bibnamefont {Tait}}, \ and\ \bibinfo {author}
  {\bibfnamefont {H.-B.}\ \bibnamefont {Yu}},\ }\href {\doibase
  10.1103/PhysRevD.82.116010} {\bibfield  {journal} {\bibinfo  {journal} {Phys.
  Rev.}\ }\textbf {\bibinfo {volume} {D82}},\ \bibinfo {pages} {116010}
  (\bibinfo {year} {2010})},\ \Eprint {http://arxiv.org/abs/1008.1783}
  {arXiv:1008.1783 [hep-ph]} \BibitemShut {NoStop}%
\bibitem [{\citenamefont {Rajaraman}\ \emph {et~al.}(2011)\citenamefont
  {Rajaraman}, \citenamefont {Shepherd}, \citenamefont {Tait},\ and\
  \citenamefont {Wijangco}}]{Rajaraman:2011wf}%
  \BibitemOpen
  \bibfield  {author} {\bibinfo {author} {\bibfnamefont {A.}~\bibnamefont
  {Rajaraman}}, \bibinfo {author} {\bibfnamefont {W.}~\bibnamefont {Shepherd}},
  \bibinfo {author} {\bibfnamefont {T.~M.~P.}\ \bibnamefont {Tait}}, \ and\
  \bibinfo {author} {\bibfnamefont {A.~M.}\ \bibnamefont {Wijangco}},\ }\href
  {\doibase 10.1103/PhysRevD.84.095013} {\bibfield  {journal} {\bibinfo
  {journal} {Phys. Rev.}\ }\textbf {\bibinfo {volume} {D84}},\ \bibinfo {pages}
  {095013} (\bibinfo {year} {2011})},\ \Eprint {http://arxiv.org/abs/1108.1196}
  {arXiv:1108.1196 [hep-ph]} \BibitemShut {NoStop}%
\bibitem [{\citenamefont {Goertz}(2014)}]{Goertz:2014qia}%
  \BibitemOpen
  \bibfield  {author} {\bibinfo {author} {\bibfnamefont {F.}~\bibnamefont
  {Goertz}},\ }\href {\doibase 10.1103/PhysRevLett.113.261803} {\bibfield
  {journal} {\bibinfo  {journal} {Phys. Rev. Lett.}\ }\textbf {\bibinfo
  {volume} {113}},\ \bibinfo {pages} {261803} (\bibinfo {year} {2014})},\
  \Eprint {http://arxiv.org/abs/1406.0102} {arXiv:1406.0102 [hep-ph]}
  \BibitemShut {NoStop}%
\bibitem [{\citenamefont {Boehm}\ \emph {et~al.}(2014)\citenamefont {Boehm},
  \citenamefont {Dolan}, \citenamefont {McCabe}, \citenamefont {Spannowsky},\
  and\ \citenamefont {Wallace}}]{Boehm:2014hva}%
  \BibitemOpen
  \bibfield  {author} {\bibinfo {author} {\bibfnamefont {C.}~\bibnamefont
  {Boehm}}, \bibinfo {author} {\bibfnamefont {M.~J.}\ \bibnamefont {Dolan}},
  \bibinfo {author} {\bibfnamefont {C.}~\bibnamefont {McCabe}}, \bibinfo
  {author} {\bibfnamefont {M.}~\bibnamefont {Spannowsky}}, \ and\ \bibinfo
  {author} {\bibfnamefont {C.~J.}\ \bibnamefont {Wallace}},\ }\href {\doibase
  10.1088/1475-7516/2014/05/009} {\bibfield  {journal} {\bibinfo  {journal}
  {JCAP}\ }\textbf {\bibinfo {volume} {1405}},\ \bibinfo {pages} {009}
  (\bibinfo {year} {2014})},\ \Eprint {http://arxiv.org/abs/1401.6458}
  {arXiv:1401.6458 [hep-ph]} \BibitemShut {NoStop}%
\bibitem [{\citenamefont {Arina}\ \emph {et~al.}(2015)\citenamefont {Arina},
  \citenamefont {Del~Nobile},\ and\ \citenamefont {Panci}}]{Arina:2014yna}%
  \BibitemOpen
  \bibfield  {author} {\bibinfo {author} {\bibfnamefont {C.}~\bibnamefont
  {Arina}}, \bibinfo {author} {\bibfnamefont {E.}~\bibnamefont {Del~Nobile}}, \
  and\ \bibinfo {author} {\bibfnamefont {P.}~\bibnamefont {Panci}},\ }\href
  {\doibase 10.1103/PhysRevLett.114.011301} {\bibfield  {journal} {\bibinfo
  {journal} {Phys. Rev. Lett.}\ }\textbf {\bibinfo {volume} {114}},\ \bibinfo
  {pages} {011301} (\bibinfo {year} {2015})},\ \Eprint
  {http://arxiv.org/abs/1406.5542} {arXiv:1406.5542 [hep-ph]} \BibitemShut
  {NoStop}%
\bibitem [{\citenamefont {Aprile}\ \emph {et~al.}(2017)\citenamefont {Aprile}
  \emph {et~al.}}]{Aprile:2017iyp}%
  \BibitemOpen
  \bibfield  {author} {\bibinfo {author} {\bibfnamefont {E.}~\bibnamefont
  {Aprile}} \emph {et~al.} (\bibinfo {collaboration} {XENON}),\ }\href
  {\doibase 10.1103/PhysRevLett.119.181301} {\bibfield  {journal} {\bibinfo
  {journal} {Phys. Rev. Lett.}\ }\textbf {\bibinfo {volume} {119}},\ \bibinfo
  {pages} {181301} (\bibinfo {year} {2017})},\ \Eprint
  {http://arxiv.org/abs/1705.06655} {arXiv:1705.06655 [astro-ph.CO]}
  \BibitemShut {NoStop}%
\bibitem [{\citenamefont {Ackermann}\ \emph {et~al.}(2015)\citenamefont
  {Ackermann} \emph {et~al.}}]{Ackermann:2015zua}%
  \BibitemOpen
  \bibfield  {author} {\bibinfo {author} {\bibfnamefont {M.}~\bibnamefont
  {Ackermann}} \emph {et~al.} (\bibinfo {collaboration} {Fermi-LAT}),\ }\href
  {\doibase 10.1103/PhysRevLett.115.231301} {\bibfield  {journal} {\bibinfo
  {journal} {Phys. Rev. Lett.}\ }\textbf {\bibinfo {volume} {115}},\ \bibinfo
  {pages} {231301} (\bibinfo {year} {2015})},\ \Eprint
  {http://arxiv.org/abs/1503.02641} {arXiv:1503.02641 [astro-ph.HE]}
  \BibitemShut {NoStop}%
\bibitem [{\citenamefont {Arcadi}\ \emph
  {et~al.}(2017{\natexlab{a}})\citenamefont {Arcadi}, \citenamefont {Dutra},
  \citenamefont {Ghosh}, \citenamefont {Lindner}, \citenamefont {Mambrini},
  \citenamefont {Pierre}, \citenamefont {Profumo},\ and\ \citenamefont
  {Queiroz}}]{Arcadi:2017kky}%
  \BibitemOpen
  \bibfield  {author} {\bibinfo {author} {\bibfnamefont {G.}~\bibnamefont
  {Arcadi}}, \bibinfo {author} {\bibfnamefont {M.}~\bibnamefont {Dutra}},
  \bibinfo {author} {\bibfnamefont {P.}~\bibnamefont {Ghosh}}, \bibinfo
  {author} {\bibfnamefont {M.}~\bibnamefont {Lindner}}, \bibinfo {author}
  {\bibfnamefont {Y.}~\bibnamefont {Mambrini}}, \bibinfo {author}
  {\bibfnamefont {M.}~\bibnamefont {Pierre}}, \bibinfo {author} {\bibfnamefont
  {S.}~\bibnamefont {Profumo}}, \ and\ \bibinfo {author} {\bibfnamefont
  {F.~S.}\ \bibnamefont {Queiroz}},\ }\href@noop {} {\  (\bibinfo {year}
  {2017}{\natexlab{a}})},\ \Eprint {http://arxiv.org/abs/1703.07364}
  {arXiv:1703.07364 [hep-ph]} \BibitemShut {NoStop}%
\bibitem [{\citenamefont {Mambrini}\ \emph {et~al.}(2016)\citenamefont
  {Mambrini}, \citenamefont {Arcadi},\ and\ \citenamefont
  {Djouadi}}]{Mambrini:2015wyu}%
  \BibitemOpen
  \bibfield  {author} {\bibinfo {author} {\bibfnamefont {Y.}~\bibnamefont
  {Mambrini}}, \bibinfo {author} {\bibfnamefont {G.}~\bibnamefont {Arcadi}}, \
  and\ \bibinfo {author} {\bibfnamefont {A.}~\bibnamefont {Djouadi}},\ }\href
  {\doibase 10.1016/j.physletb.2016.02.049} {\bibfield  {journal} {\bibinfo
  {journal} {Phys. Lett.}\ }\textbf {\bibinfo {volume} {B755}},\ \bibinfo
  {pages} {426} (\bibinfo {year} {2016})},\ \Eprint
  {http://arxiv.org/abs/1512.04913} {arXiv:1512.04913 [hep-ph]} \BibitemShut
  {NoStop}%
\bibitem [{\citenamefont {Arcadi}\ \emph
  {et~al.}(2017{\natexlab{b}})\citenamefont {Arcadi}, \citenamefont {Lindner},
  \citenamefont {Queiroz}, \citenamefont {Rodejohann},\ and\ \citenamefont
  {Vogl}}]{Arcadi:2017wqi}%
  \BibitemOpen
  \bibfield  {author} {\bibinfo {author} {\bibfnamefont {G.}~\bibnamefont
  {Arcadi}}, \bibinfo {author} {\bibfnamefont {M.}~\bibnamefont {Lindner}},
  \bibinfo {author} {\bibfnamefont {F.~S.}\ \bibnamefont {Queiroz}}, \bibinfo
  {author} {\bibfnamefont {W.}~\bibnamefont {Rodejohann}}, \ and\ \bibinfo
  {author} {\bibfnamefont {S.}~\bibnamefont {Vogl}},\ }\href@noop {} {\
  (\bibinfo {year} {2017}{\natexlab{b}})},\ \Eprint
  {http://arxiv.org/abs/1711.02110} {arXiv:1711.02110 [hep-ph]} \BibitemShut
  {NoStop}%
\bibitem [{\citenamefont {Bauer}\ \emph
  {et~al.}(2017{\natexlab{b}})\citenamefont {Bauer}, \citenamefont {Klassen},\
  and\ \citenamefont {Tenorth}}]{Bauer:2017fsw}%
  \BibitemOpen
  \bibfield  {author} {\bibinfo {author} {\bibfnamefont {M.}~\bibnamefont
  {Bauer}}, \bibinfo {author} {\bibfnamefont {M.}~\bibnamefont {Klassen}}, \
  and\ \bibinfo {author} {\bibfnamefont {V.}~\bibnamefont {Tenorth}},\
  }\href@noop {} {\  (\bibinfo {year} {2017}{\natexlab{b}})},\ \Eprint
  {http://arxiv.org/abs/1712.06597} {arXiv:1712.06597 [hep-ph]} \BibitemShut
  {NoStop}%
\bibitem [{\citenamefont {Alanne}\ \emph {et~al.}(2014)\citenamefont {Alanne},
  \citenamefont {Tuominen},\ and\ \citenamefont {Vaskonen}}]{Alanne:2014bra}%
  \BibitemOpen
  \bibfield  {author} {\bibinfo {author} {\bibfnamefont {T.}~\bibnamefont
  {Alanne}}, \bibinfo {author} {\bibfnamefont {K.}~\bibnamefont {Tuominen}}, \
  and\ \bibinfo {author} {\bibfnamefont {V.}~\bibnamefont {Vaskonen}},\ }\href
  {\doibase 10.1016/j.nuclphysb.2014.11.001} {\bibfield  {journal} {\bibinfo
  {journal} {Nucl. Phys.}\ }\textbf {\bibinfo {volume} {B889}},\ \bibinfo
  {pages} {692} (\bibinfo {year} {2014})},\ \Eprint
  {http://arxiv.org/abs/1407.0688} {arXiv:1407.0688 [hep-ph]} \BibitemShut
  {NoStop}%
\bibitem [{\citenamefont {Hill}\ and\ \citenamefont
  {Solon}(2015)}]{Hill:2014yxa}%
  \BibitemOpen
  \bibfield  {author} {\bibinfo {author} {\bibfnamefont {R.~J.}\ \bibnamefont
  {Hill}}\ and\ \bibinfo {author} {\bibfnamefont {M.~P.}\ \bibnamefont
  {Solon}},\ }\href {\doibase 10.1103/PhysRevD.91.043505} {\bibfield  {journal}
  {\bibinfo  {journal} {Phys. Rev.}\ }\textbf {\bibinfo {volume} {D91}},\
  \bibinfo {pages} {043505} (\bibinfo {year} {2015})},\ \Eprint
  {http://arxiv.org/abs/1409.8290} {arXiv:1409.8290 [hep-ph]} \BibitemShut
  {NoStop}%
\bibitem [{\citenamefont {Bishara}\ \emph
  {et~al.}(2017{\natexlab{a}})\citenamefont {Bishara}, \citenamefont {Brod},
  \citenamefont {Grinstein},\ and\ \citenamefont {Zupan}}]{Bishara:2017pfq}%
  \BibitemOpen
  \bibfield  {author} {\bibinfo {author} {\bibfnamefont {F.}~\bibnamefont
  {Bishara}}, \bibinfo {author} {\bibfnamefont {J.}~\bibnamefont {Brod}},
  \bibinfo {author} {\bibfnamefont {B.}~\bibnamefont {Grinstein}}, \ and\
  \bibinfo {author} {\bibfnamefont {J.}~\bibnamefont {Zupan}},\ }\href
  {\doibase 10.1007/JHEP11(2017)059} {\bibfield  {journal} {\bibinfo  {journal}
  {JHEP}\ }\textbf {\bibinfo {volume} {11}},\ \bibinfo {pages} {059} (\bibinfo
  {year} {2017}{\natexlab{a}})},\ \Eprint {http://arxiv.org/abs/1707.06998}
  {arXiv:1707.06998 [hep-ph]} \BibitemShut {NoStop}%
\bibitem [{\citenamefont {Bishara}\ \emph
  {et~al.}(2017{\natexlab{b}})\citenamefont {Bishara}, \citenamefont {Brod},
  \citenamefont {Grinstein},\ and\ \citenamefont {Zupan}}]{Bishara:2017nnn}%
  \BibitemOpen
  \bibfield  {author} {\bibinfo {author} {\bibfnamefont {F.}~\bibnamefont
  {Bishara}}, \bibinfo {author} {\bibfnamefont {J.}~\bibnamefont {Brod}},
  \bibinfo {author} {\bibfnamefont {B.}~\bibnamefont {Grinstein}}, \ and\
  \bibinfo {author} {\bibfnamefont {J.}~\bibnamefont {Zupan}},\ }\href@noop {}
  {\  (\bibinfo {year} {2017}{\natexlab{b}})},\ \Eprint
  {http://arxiv.org/abs/1708.02678} {arXiv:1708.02678 [hep-ph]} \BibitemShut
  {NoStop}%
\bibitem [{\citenamefont {Alwall}\ \emph {et~al.}(2014)\citenamefont {Alwall},
  \citenamefont {Frederix}, \citenamefont {Frixione}, \citenamefont {Hirschi},
  \citenamefont {Maltoni}, \citenamefont {Mattelaer}, \citenamefont {Shao},
  \citenamefont {Stelzer}, \citenamefont {Torrielli},\ and\ \citenamefont
  {Zaro}}]{Alwall:2014hca}%
  \BibitemOpen
  \bibfield  {author} {\bibinfo {author} {\bibfnamefont {J.}~\bibnamefont
  {Alwall}}, \bibinfo {author} {\bibfnamefont {R.}~\bibnamefont {Frederix}},
  \bibinfo {author} {\bibfnamefont {S.}~\bibnamefont {Frixione}}, \bibinfo
  {author} {\bibfnamefont {V.}~\bibnamefont {Hirschi}}, \bibinfo {author}
  {\bibfnamefont {F.}~\bibnamefont {Maltoni}}, \bibinfo {author} {\bibfnamefont
  {O.}~\bibnamefont {Mattelaer}}, \bibinfo {author} {\bibfnamefont {H.~S.}\
  \bibnamefont {Shao}}, \bibinfo {author} {\bibfnamefont {T.}~\bibnamefont
  {Stelzer}}, \bibinfo {author} {\bibfnamefont {P.}~\bibnamefont {Torrielli}},
  \ and\ \bibinfo {author} {\bibfnamefont {M.}~\bibnamefont {Zaro}},\ }\href
  {\doibase 10.1007/JHEP07(2014)079} {\bibfield  {journal} {\bibinfo  {journal}
  {JHEP}\ }\textbf {\bibinfo {volume} {07}},\ \bibinfo {pages} {079} (\bibinfo
  {year} {2014})},\ \Eprint {http://arxiv.org/abs/1405.0301} {arXiv:1405.0301
  [hep-ph]} \BibitemShut {NoStop}%
\bibitem [{\citenamefont {McDonald}(1994{\natexlab{a}})}]{McDonald:1993ex}%
  \BibitemOpen
  \bibfield  {author} {\bibinfo {author} {\bibfnamefont {J.}~\bibnamefont
  {McDonald}},\ }\href {\doibase 10.1103/PhysRevD.50.3637} {\bibfield
  {journal} {\bibinfo  {journal} {Phys. Rev.}\ }\textbf {\bibinfo {volume}
  {D50}},\ \bibinfo {pages} {3637} (\bibinfo {year} {1994}{\natexlab{a}})},\
  \Eprint {http://arxiv.org/abs/hep-ph/0702143} {arXiv:hep-ph/0702143 [HEP-PH]}
  \BibitemShut {NoStop}%
\bibitem [{\citenamefont {Burgess}\ \emph {et~al.}(2001)\citenamefont
  {Burgess}, \citenamefont {Pospelov},\ and\ \citenamefont {ter
  Veldhuis}}]{Burgess:2000yq}%
  \BibitemOpen
  \bibfield  {author} {\bibinfo {author} {\bibfnamefont {C.~P.}\ \bibnamefont
  {Burgess}}, \bibinfo {author} {\bibfnamefont {M.}~\bibnamefont {Pospelov}}, \
  and\ \bibinfo {author} {\bibfnamefont {T.}~\bibnamefont {ter Veldhuis}},\
  }\href {\doibase 10.1016/S0550-3213(01)00513-2} {\bibfield  {journal}
  {\bibinfo  {journal} {Nucl. Phys.}\ }\textbf {\bibinfo {volume} {B619}},\
  \bibinfo {pages} {709} (\bibinfo {year} {2001})},\ \Eprint
  {http://arxiv.org/abs/hep-ph/0011335} {arXiv:hep-ph/0011335 [hep-ph]}
  \BibitemShut {NoStop}%
\bibitem [{\citenamefont {Cline}\ \emph {et~al.}(2013)\citenamefont {Cline},
  \citenamefont {Kainulainen}, \citenamefont {Scott},\ and\ \citenamefont
  {Weniger}}]{Cline:2013gha}%
  \BibitemOpen
  \bibfield  {author} {\bibinfo {author} {\bibfnamefont {J.~M.}\ \bibnamefont
  {Cline}}, \bibinfo {author} {\bibfnamefont {K.}~\bibnamefont {Kainulainen}},
  \bibinfo {author} {\bibfnamefont {P.}~\bibnamefont {Scott}}, \ and\ \bibinfo
  {author} {\bibfnamefont {C.}~\bibnamefont {Weniger}},\ }\href {\doibase
  10.1103/PhysRevD.92.039906, 10.1103/PhysRevD.88.055025} {\bibfield  {journal}
  {\bibinfo  {journal} {Phys. Rev.}\ }\textbf {\bibinfo {volume} {D88}},\
  \bibinfo {pages} {055025} (\bibinfo {year} {2013})},\ \bibinfo {note}
  {[Erratum: Phys. Rev.D92,no.3,039906(2015)]},\ \Eprint
  {http://arxiv.org/abs/1306.4710} {arXiv:1306.4710 [hep-ph]} \BibitemShut
  {NoStop}%
\bibitem [{\citenamefont {Lopez-Honorez}\ \emph {et~al.}(2012)\citenamefont
  {Lopez-Honorez}, \citenamefont {Schwetz},\ and\ \citenamefont
  {Zupan}}]{LopezHonorez:2012kv}%
  \BibitemOpen
  \bibfield  {author} {\bibinfo {author} {\bibfnamefont {L.}~\bibnamefont
  {Lopez-Honorez}}, \bibinfo {author} {\bibfnamefont {T.}~\bibnamefont
  {Schwetz}}, \ and\ \bibinfo {author} {\bibfnamefont {J.}~\bibnamefont
  {Zupan}},\ }\href {\doibase 10.1016/j.physletb.2012.07.017} {\bibfield
  {journal} {\bibinfo  {journal} {Phys. Lett.}\ }\textbf {\bibinfo {volume}
  {B716}},\ \bibinfo {pages} {179} (\bibinfo {year} {2012})},\ \Eprint
  {http://arxiv.org/abs/1203.2064} {arXiv:1203.2064 [hep-ph]} \BibitemShut
  {NoStop}%
\bibitem [{\citenamefont {Cline}\ and\ \citenamefont
  {Kainulainen}(2013)}]{Cline:2012hg}%
  \BibitemOpen
  \bibfield  {author} {\bibinfo {author} {\bibfnamefont {J.~M.}\ \bibnamefont
  {Cline}}\ and\ \bibinfo {author} {\bibfnamefont {K.}~\bibnamefont
  {Kainulainen}},\ }\href {\doibase 10.1088/1475-7516/2013/01/012} {\bibfield
  {journal} {\bibinfo  {journal} {JCAP}\ }\textbf {\bibinfo {volume} {1301}},\
  \bibinfo {pages} {012} (\bibinfo {year} {2013})},\ \Eprint
  {http://arxiv.org/abs/1210.4196} {arXiv:1210.4196 [hep-ph]} \BibitemShut
  {NoStop}%
\bibitem [{\citenamefont {Kang}\ \emph {et~al.}(2016)\citenamefont {Kang},
  \citenamefont {Ko},\ and\ \citenamefont {Li}}]{Kang:2015nga}%
  \BibitemOpen
  \bibfield  {author} {\bibinfo {author} {\bibfnamefont {Z.}~\bibnamefont
  {Kang}}, \bibinfo {author} {\bibfnamefont {P.}~\bibnamefont {Ko}}, \ and\
  \bibinfo {author} {\bibfnamefont {J.}~\bibnamefont {Li}},\ }\href {\doibase
  10.1103/PhysRevLett.116.131801} {\bibfield  {journal} {\bibinfo  {journal}
  {Phys. Rev. Lett.}\ }\textbf {\bibinfo {volume} {116}},\ \bibinfo {pages}
  {131801} (\bibinfo {year} {2016})},\ \Eprint
  {http://arxiv.org/abs/1504.04128} {arXiv:1504.04128 [hep-ph]} \BibitemShut
  {NoStop}%
\bibitem [{\citenamefont {Kajantie}\ \emph {et~al.}(1996)\citenamefont
  {Kajantie}, \citenamefont {Laine}, \citenamefont {Rummukainen},\ and\
  \citenamefont {Shaposhnikov}}]{Kajantie:1996mn}%
  \BibitemOpen
  \bibfield  {author} {\bibinfo {author} {\bibfnamefont {K.}~\bibnamefont
  {Kajantie}}, \bibinfo {author} {\bibfnamefont {M.}~\bibnamefont {Laine}},
  \bibinfo {author} {\bibfnamefont {K.}~\bibnamefont {Rummukainen}}, \ and\
  \bibinfo {author} {\bibfnamefont {M.~E.}\ \bibnamefont {Shaposhnikov}},\
  }\href {\doibase 10.1103/PhysRevLett.77.2887} {\bibfield  {journal} {\bibinfo
   {journal} {Phys. Rev. Lett.}\ }\textbf {\bibinfo {volume} {77}},\ \bibinfo
  {pages} {2887} (\bibinfo {year} {1996})},\ \Eprint
  {http://arxiv.org/abs/hep-ph/9605288} {arXiv:hep-ph/9605288 [hep-ph]}
  \BibitemShut {NoStop}%
\bibitem [{\citenamefont {Rummukainen}\ \emph {et~al.}(1998)\citenamefont
  {Rummukainen}, \citenamefont {Tsypin}, \citenamefont {Kajantie},
  \citenamefont {Laine},\ and\ \citenamefont
  {Shaposhnikov}}]{Rummukainen:1998as}%
  \BibitemOpen
  \bibfield  {author} {\bibinfo {author} {\bibfnamefont {K.}~\bibnamefont
  {Rummukainen}}, \bibinfo {author} {\bibfnamefont {M.}~\bibnamefont {Tsypin}},
  \bibinfo {author} {\bibfnamefont {K.}~\bibnamefont {Kajantie}}, \bibinfo
  {author} {\bibfnamefont {M.}~\bibnamefont {Laine}}, \ and\ \bibinfo {author}
  {\bibfnamefont {M.~E.}\ \bibnamefont {Shaposhnikov}},\ }\href {\doibase
  10.1016/S0550-3213(98)00494-5} {\bibfield  {journal} {\bibinfo  {journal}
  {Nucl. Phys.}\ }\textbf {\bibinfo {volume} {B532}},\ \bibinfo {pages} {283}
  (\bibinfo {year} {1998})},\ \Eprint {http://arxiv.org/abs/hep-lat/9805013}
  {arXiv:hep-lat/9805013 [hep-lat]} \BibitemShut {NoStop}%
\bibitem [{\citenamefont {Anderson}\ and\ \citenamefont
  {Hall}(1992)}]{Anderson:1991zb}%
  \BibitemOpen
  \bibfield  {author} {\bibinfo {author} {\bibfnamefont {G.~W.}\ \bibnamefont
  {Anderson}}\ and\ \bibinfo {author} {\bibfnamefont {L.~J.}\ \bibnamefont
  {Hall}},\ }\href {\doibase 10.1103/PhysRevD.45.2685} {\bibfield  {journal}
  {\bibinfo  {journal} {Phys. Rev.}\ }\textbf {\bibinfo {volume} {D45}},\
  \bibinfo {pages} {2685} (\bibinfo {year} {1992})}\BibitemShut {NoStop}%
\bibitem [{\citenamefont {Espinosa}\ and\ \citenamefont
  {Quiros}(1993)}]{Espinosa:1993bs}%
  \BibitemOpen
  \bibfield  {author} {\bibinfo {author} {\bibfnamefont {J.~R.}\ \bibnamefont
  {Espinosa}}\ and\ \bibinfo {author} {\bibfnamefont {M.}~\bibnamefont
  {Quiros}},\ }\href {\doibase 10.1016/0370-2693(93)91111-Y} {\bibfield
  {journal} {\bibinfo  {journal} {Phys. Lett.}\ }\textbf {\bibinfo {volume}
  {B305}},\ \bibinfo {pages} {98} (\bibinfo {year} {1993})},\ \Eprint
  {http://arxiv.org/abs/hep-ph/9301285} {arXiv:hep-ph/9301285 [hep-ph]}
  \BibitemShut {NoStop}%
\bibitem [{\citenamefont {Choi}\ and\ \citenamefont
  {Volkas}(1993)}]{Choi:1993cv}%
  \BibitemOpen
  \bibfield  {author} {\bibinfo {author} {\bibfnamefont {J.}~\bibnamefont
  {Choi}}\ and\ \bibinfo {author} {\bibfnamefont {R.~R.}\ \bibnamefont
  {Volkas}},\ }\href {\doibase 10.1016/0370-2693(93)91013-D} {\bibfield
  {journal} {\bibinfo  {journal} {Phys. Lett.}\ }\textbf {\bibinfo {volume}
  {B317}},\ \bibinfo {pages} {385} (\bibinfo {year} {1993})},\ \Eprint
  {http://arxiv.org/abs/hep-ph/9308234} {arXiv:hep-ph/9308234 [hep-ph]}
  \BibitemShut {NoStop}%
\bibitem [{\citenamefont {McDonald}(1994{\natexlab{b}})}]{McDonald:1993ey}%
  \BibitemOpen
  \bibfield  {author} {\bibinfo {author} {\bibfnamefont {J.}~\bibnamefont
  {McDonald}},\ }\href {\doibase 10.1016/0370-2693(94)91229-7} {\bibfield
  {journal} {\bibinfo  {journal} {Phys. Lett.}\ }\textbf {\bibinfo {volume}
  {B323}},\ \bibinfo {pages} {339} (\bibinfo {year}
  {1994}{\natexlab{b}})}\BibitemShut {NoStop}%
\end{thebibliography}%

\end{document}